\documentclass[prl,notitlepage,twocolumn]{revtex4}

\usepackage{graphicx}
\usepackage{amsmath}
\usepackage{amssymb}
\usepackage{amsfonts}
\usepackage{color}
\usepackage{xspace}
\usepackage{subfigure}

\newcommand{\pdmit}{Pd(dmit)$_2$\xspace}
\newcommand{\moly}{Mo$_3$S$_7$(dmit)$_3$\xspace}

\newcommand{\dmitSL}{EtMe$_3$Sb[{\pdmit}]$_2$}

\begin{document}

\title{Deriving \textit{ab initio} model Hamiltonians for molecular crystals}

\author{A. C. Jacko}\affiliation{School of Mathematics and Physics, The University of Queensland, Brisbane, Queensland, 4072, Australia}

\begin{abstract}
Developing realistic and precise models of the electronic properties of organic molecular crystals is crucial for understanding the full range of strongly correlated phases that they exhibit. By using \textit{ab initio} model construction methods, one can obtain unbiased non-interacting models of such systems from density functional theory, upon which one can base further (many-body) models. 
We will discuss the utility and advantages of \textit{ab initio} model construction using Wannier orbitals. We will briefly review the approach, and then explain why it is so well suited to molecular crystals in particular. 
We discuss the \textit{ab initio} construction of both non-interacting and interacting Hamitonians, and highlight recent examples where such first principles models lead to importantly different results than fitted models.
\end{abstract}

\maketitle
\pagebreak
\tableofcontents

\section{Background}
\subsection{Molecular Crystals}
Molecular crystals, ordered periodic arrays of molecules, are known to exhibit a wide range of quantum mechanical phenomena, including unconventional superconductivity, quantum criticality, frustrated anti-ferromagnetism, 
and quantum spin liquid behaviour \cite{chaikin98, powell11, jacko13dmit, dressel07, seo04, brown15, jacko13tmttf, seo15}. In some cases many of these phases can be found in a single material by tuning an external parameter, such as pressure or magnetic field. Often, one can control which phase is expressed by making subtle physical or chemical changes to the molecules \cite{seo04, dressel07, jacko13tmttf, seo15,brown15}. Along with the flexibility of the interactions within individual molecules, molecular crystals also have a range of intermolecular interactions. It is the subtle competition between the many intra- and inter-molecular interaction energies that brings the wide variety of phases seen in experiments so close together. These crystals tend to have a low effective dimension, and this likely contributes to the close competition between the phases \cite{seo04, dressel07,brown15,coldea10}.

Molecular crystals are an exciting testing ground for finding and understanding new emergent states of matter. They often display competition between multiple emergent strongly correlated ground states \cite{dressel07,powell11}. This, and the flexibility of organic chemistry, means that the emergent physics is often tuneable by subtle chemical and physical modifications (making slight variations around a core motif) \cite{seo04, dressel07, jacko13tmttf, seo15,brown15}. 
In one notable case, a superconducting state in an organic molecular crystal is destroyed                                by substituting some hydrogen atoms for deuterium, its heavy isotope \cite{taniguchi99}. 
That this extremely subtle change can have such profound consequences is both exciting and intimidating. On one hand, it presents the inviting prospect of creating strongly correlated materials with technologically desireable properties; on the other, the level of detail required to correctly predict the phase of a material can be substatial.

\begin{figure}
\begin{center}
 \includegraphics[width=0.96\columnwidth]{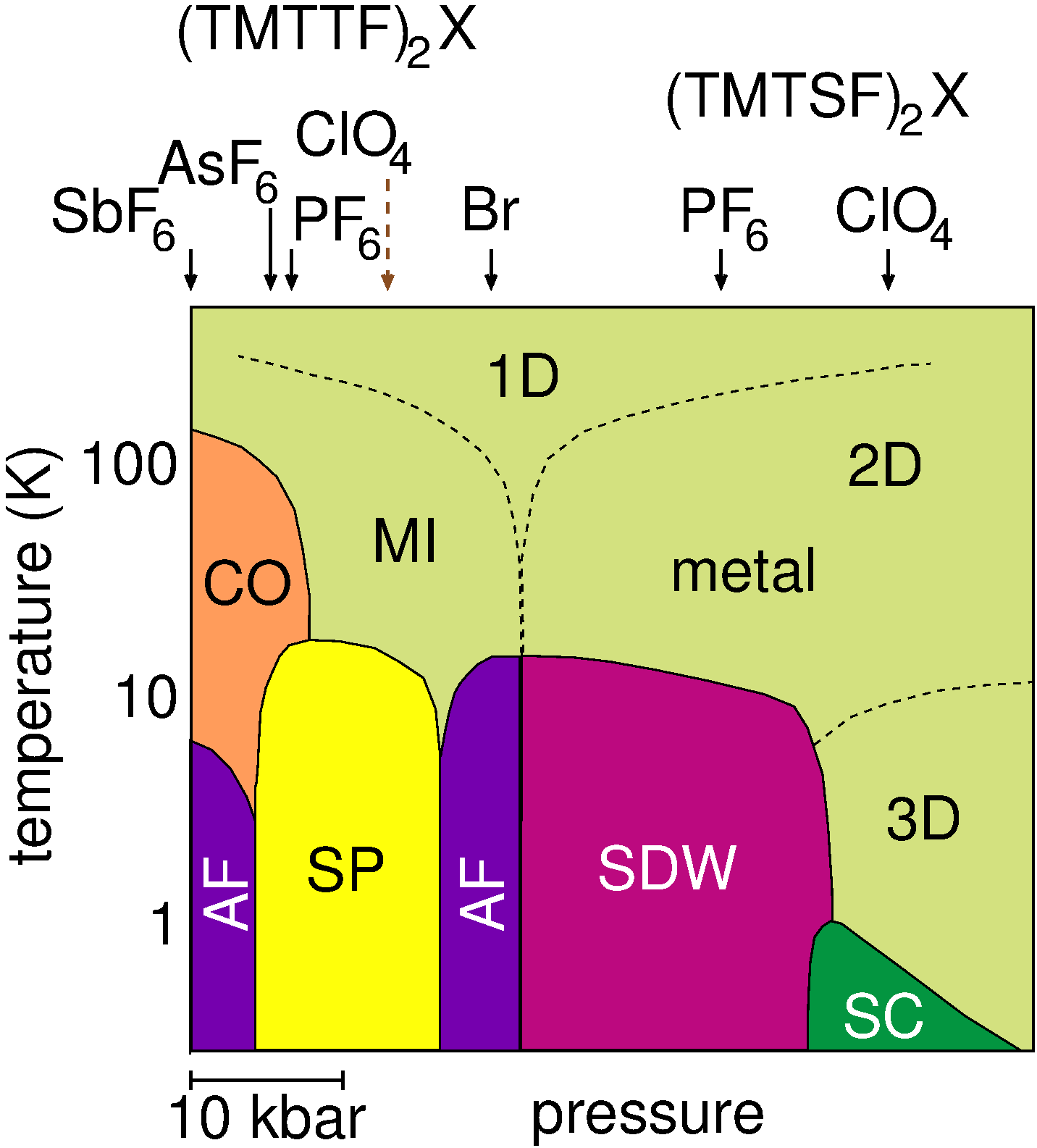} 
 \caption{Temperature-pressure phase diagram for the Fabre (TMTTF) and Bechgaard (TMTSF) charge transfer salts \cite{jerome91,jacko13tmttf}, highlighting the many accessible strongly-correlated phases. The phases shown are antiferromagnetic (AF), charge ordered (CO), Mott insulating (MI), spin Peierls (SP), spin density wave (SDW), superconducting (SC) and one dimensional (1D), 2D, \& 3D metals. The ambient pressure position for each salt is indicated with an arrow above the diagram. }\label{fig:fabrephase}
\end{center}
\end{figure}

Fig. \ref{fig:fabrephase} shows the phase diagram for a family of organic crystals called Fabre salts [salts of TMTTF (shown in Fig. \ref{fig:molvscrystal}) and an anion], which can be tuned through many different phases by applying physical pressure, or by slightly changing their anions, often thought of as a ‘chemical pressure’. As the size of the anion decreases, the TMTTF molecules pack closer together, as they would under the application of physical pressure. This range of accessible phases indicates that the many competing interactions are very close in energy. The many competing energy scales in molecular crystals gives us access to phases with various interesting physical properties.

\subsection{Predictive versus postdictive modeling}

There is an important philosophical point to be made here. The goal of science should be to make predictions about the nature of nature, and then to test those predictions. In the case of building models of materials, what is typically applied is a \textit{postdictive} approach; one `knows' that to have a model with the observed behaviour, it should be \textit{this} lattice with \textit{that} type of interaction (e.g. Heisenberg model on a triangular lattice, extended Hubbard model on a square lattice, and so on). 
Thus, there is limited information gained \textbf{about the system}, whether one finds the behaviour one was searching for in such a model or not. One one hand, one chose the model phenomenologically, so finding the correct phenomenology is not profound. On the other, \textbf{not} finding the expected behaviour could be due to any number of reasons from the profound to the trite.

Postdictive approaches can be useful, but one must go beyond them to gain a deeper understanding of the important commonalities and differences within a class of materials. For example, such an approach does not show much promise for describing all of the multitude of phases seen in the Fabre salts.
What one ideally would like is a systematic way of constructing an effective many-body Hamiltonian from first principles. Constructing the non-interacting part from first principles is currently emminantly possible:
By producing localised `Wannier' orbitals (discussed in more detail later), one can use the results of density functional theory (DFT) to construct a tight-binding lattice without first assuming its form. Here we give a brief overview of the theoretical development of the concept and practical details of using Wannier orbitals. For a much more complete and mathematical account, turn to the excellent review of Marzari \textit{et al.} \cite{marzari12}.

On a related note, it is worth commenting on the distinction between `\textit{ab initio}' and `first principles'; \textit{ab initio} implies no empirical input, just calculations on the grounds of the many-electron Schr{\"o}dinger equation using the fundamental constants of nature such as Planck's constant, the charge of the electron, etc. On the other hand, first principles allows for empirical parameters. Both density functional theory and Wannier orbital construction are in principle \textit{ab initio}, however particular implementations tend to include empirical parameterisations (specific density functionals, for example) that are properly considered first principles rather than \textit{ab initio}.

\subsection{Development of Wannier Orbitals}

In 1937 Gregory Wannier introduced the idea of constructing localised sets of wavefunctions by fourier transforming Bloch states \cite{wannier37}. For a Bloch wavefunction for band $n$, $\psi_n(\mathbf{k},\mathbf{r})$, the corresponding Wannier function for band $n$ is
\begin{equation}\label{eq:wannier}
\Phi_{n,\mathbf{R}}(\mathbf{r}) = \int_{FBZ} d^3 \mathbf{k} e^{-i\mathbf{k}.\mathbf{R}}\psi_n(\mathbf{k},\mathbf{r}),
\end{equation}
where $\mathbf{R}$ is any combination of the crystal lattice vectors with integer prefactors, $\Phi_{\mathbf{R}}(\mathbf{r})$ is localised in the unit cell located at $\mathbf{R}$, and the integral runs over the first Brillouin zone (FBZ). 
These new wavefunctions have the advantages of atomic orbitals (such as locality) while enforcing orthogonality. This allows one to treat localised excitations of individual electrons in metallic materials on the same footing as the `bulk' electrons (the delocalised Bloch states).

By the 1950's these wavefunctions were widely known as Wannier functions, and of great use in understanding the physics of excitations in crystals. In 1953, George Koster introduced two new methods for defining Wannier functions without first having to solve the Schr\"{o}dinger equation, and allowing one to use these orbitals to compute energy bands in crystals \cite{koster53}.

Walter Kohn put Wannier functions on a rigourous analytical grounding in 1959, showing that one can always find a unique, real, symmetry-preserving, and exponentially localised functions for a given single band \cite{kohn59}. In this work he showed (although not in so many words) that Wannier orbitals were the ideal basis for the recently-developed tight-binding method (closely related to the H\"uckel method used in chemistry) \cite{huckel31,slater54}. (Kohn continued working on Wannier orbitals, and in the mid-90's used the locality of Wannier functions, and the consequence that their interactions should decay exponentially, to propose a density functional theory method that scales linearly with the number of atoms \cite{kohn93, kohn95}.) Jacques Des Cloizeaux further expanded the mathematical grounding of Wannier functions, and identified what would later become known as the disentangling problem: if bands overlap, it is difficult to construct Wannier functions for just one of those bands (requiring one to `disentangle' the target band from the other bands it crosses) \cite{descloizeaux64}. This remains a general challenge in using Wannier functions to this day \cite{marzari12}. Due to the practical difficulty of the disentangling problem, and the extra indeterminancy introduced in the disentangling proceedure, Wanner functions were not of signficant help in computational electronic structure theory until the 90's, when approaches based on density functional theory were introduced \cite{marzari97}.

\subsection{Wannier orbitals in Density Functional Theory}\label{sec:wodft}
The key breakthrough in the application of Wannier orbitals occured when Nicola Marzari and David Vanderbilt formulated a generalised approach for generating maximally localised Wannier functions for the case of multiple bands \cite{marzari97}. Not only that, they also described a numerical algorithm to produce such orbitals based on Bloch functions sampled on a mesh of points in $k$-space, such as would be the output of a typical DFT code. This allowed for computations of Wannier orbitals in realistic, non-trivial cases. They also suggested the approach of using Wannier functions to construct effective model Hamiltonians for strongly correlated electron systems. 

A few years later, Marzari and Vanderbilt, along with Ivo Souza, extended their original approach to allow for entangled bands. Together they introduced an efficient disentangling methodology \cite{souza01} that requires no additional information over a usual Wannier construction, just one additional assumption. That assumption is that the `character' of the Wannier orbitals (the contributions from particular basis functions) should vary as smoothly and slowly as possible; this is enforced via minimising the change in character across the Brillouin zone.

These works laid the foundation for the wide-spread computation of Wannier orbitals in DFT codes. In 2008, the code \texttt{wannier90} was released \cite{mostofi08}. Developed by Arash Mostofi, Jonathan Yates, Young-Su Lee, along with Souza, Vanderbilt and Marzari, this code is now widely used, designed to interface with any DFT code to produce Wannier orbitals. 
It is now used in Wannier orbital construction in FPLO \cite{koepernik99}, WIEN2k \cite{kunes10}, Quantum ESPRESSO \cite{giannozzi09}, ABINIT \cite{gonze09}, and Fleur \cite{freimuth08}, to list just a few of the more popular DFT codes for crystals.

\section{Separation of energy scales in Molecular Crystals}
Here we discuss the separation of energy scales that commonly occurs in molecular crystals and how this aids the construction of a minimal set of Wannier orbitals.
Molecular crystals tend to have a separation of energy scales in their non-interacting states, while there is competition amongst many possible strongly correlated ground states in the full many-body treatment.  Despite the advances made in disentangling procedures, it remains a highly challenging task to produce high quality, reliable Wannier orbitals from entangled bands. This is particularly important if one is concerned about capturing the fine features of the electronic structure, which can have significant effects on the many-body state, as I will discuss explicitly in the case of crystals based on {\pdmit}.
Molecular crystals provide an exciting playground for applying Wannier orbital based techniques to their greatest potential. because one can bypass the difficulty and ambiguity of disentangling procedures, one can determine the significance of the fine features of the electronic structure in determining the rich phase diagrams of these materials.

\begin{figure}
\begin{center}
 \includegraphics[width=0.95\columnwidth]{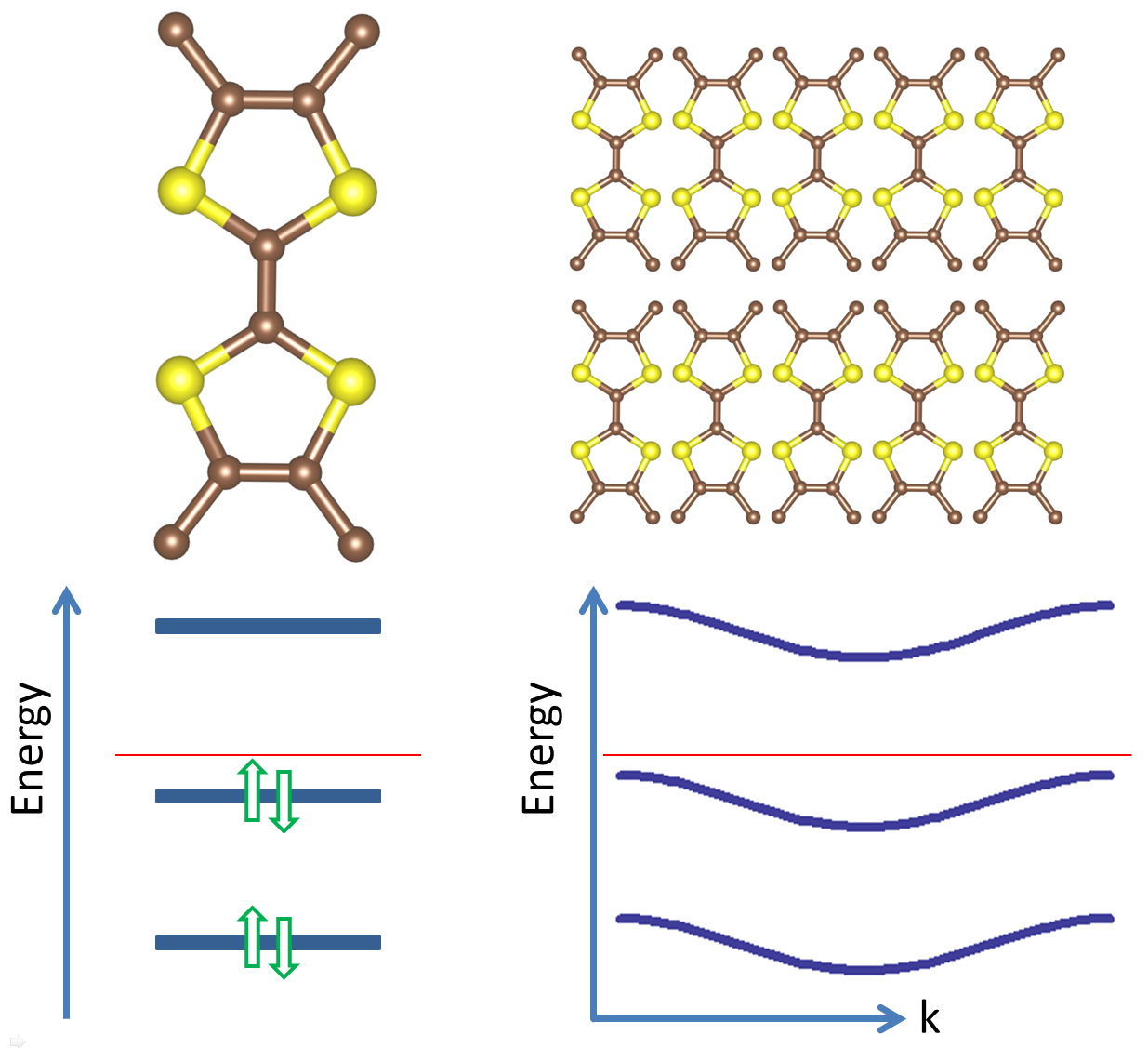} 
 \caption{Discrete energy levels of a single molecule (left) (here, TMTTF \textit{sans} hydrogens), and continuum band states of the resulting molecular crystal (right). The interatomic interactions within the molecule set the energy scale for the molecular orbitals, while the inter-molecular coupling determines the width of the bands. In typical cases these energy scale are quite different (as illustrated).}\label{fig:molvscrystal}
\end{center}
\end{figure}

The separation of energy scales in molecular crystals is straight-forward to understand: the molecules are held together by (strong) covalent bonds, while the crystal is held together by much weaker intermolecular forces; van der Waals, $\pi$-stacking, and hydrogen bonding. The strong forces within a molecule produce a set of well spaced molecular orbitals (MOs), and these orbitals weakly couple between molecules, as illustrated in Fig. \ref{fig:molvscrystal}, producing bands that are narrow on the scale of the MO energy gaps.

\begin{figure}
\begin{center}
 \includegraphics[width=0.95\columnwidth]{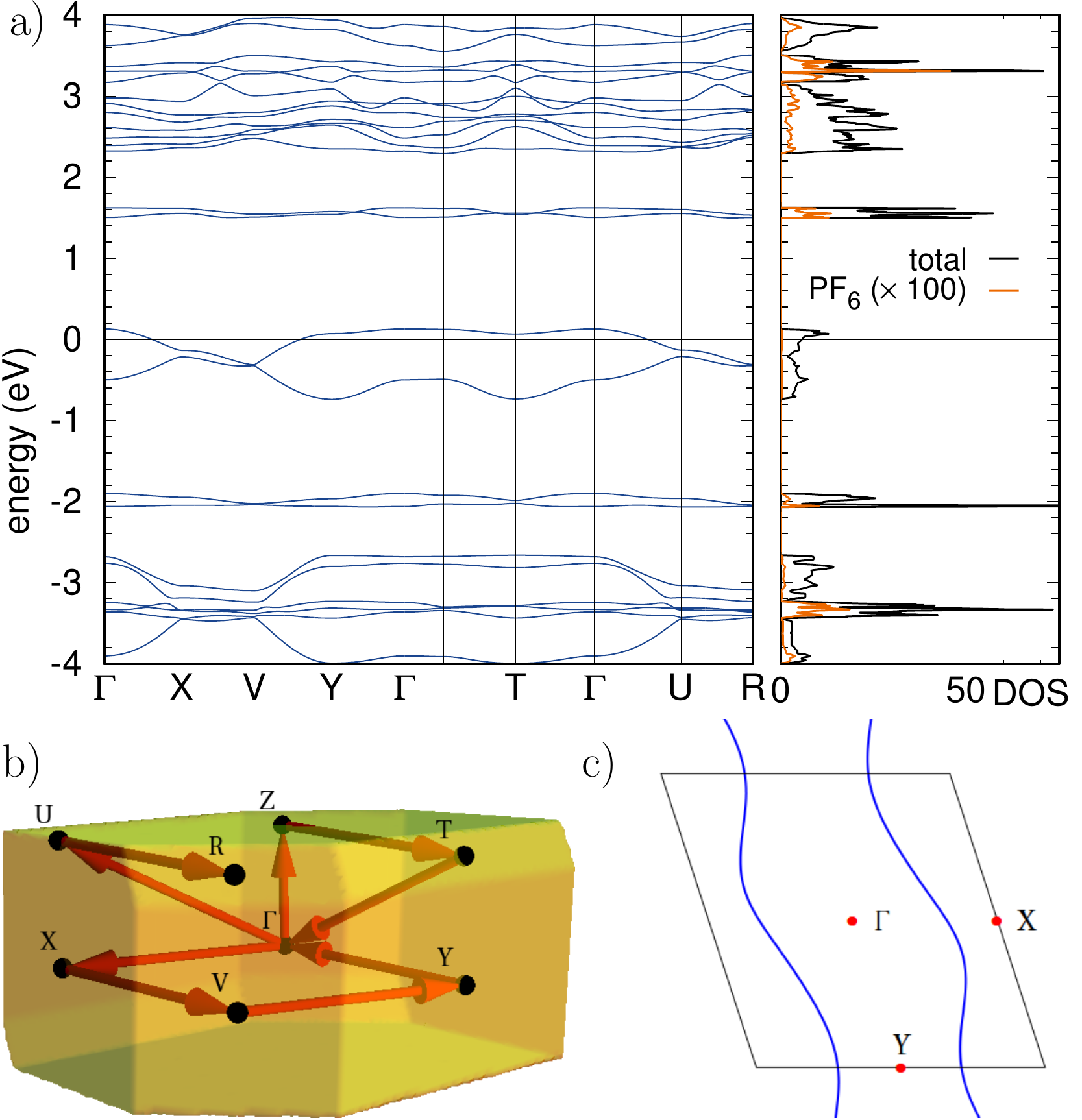} 
 \caption{Electronic properties 
of (TMTTF)$_2$PF$_6$  at $T=4$~K, reporduced from \cite{jacko13tmttf}. a) Band structure and density of states,
b) path through $k$-space, c) Fermi surface in the $k_z = 0$ plane.  The total density of states is shown with the solid black line, and the
  partial density of states of the anions ($\times$100) is shown in dashed orange. The partial density of states shows
  that the two bands at the Fermi level are nearly purely TMTTF, with large energy gaps on either side, demonstrating the separation of energy scales.}\label{fig:bsdos}
\end{center}
\end{figure}

We can understand this more concretely by considering a toy example: a 1D, two orbital tight-binding model. For simplicity we consider a chain of spinless fermions, with orbitals $a$ and $b$ on each site $i$, with nearest neighbour interactions. The Hamiltonian is
\begin{equation}
\hat{H} = \sum_{i} \frac{\Delta}{2} (\hat{n}_{a,i}-\hat{n}_{b,i}) + \sum_{\alpha=a,b} t_\alpha \hat{c}_{\alpha,i}^\dagger \hat{c}_{\alpha,i+1}  + h.c.
\end{equation}
where $\Delta$ is the energy difference between the orbitals, and $t_\alpha$ is the inter-site hopping for orbital $\alpha$. 
In the case of molecular crystals, the orbitals are molecular orbitals of single molecules. The energy difference between the molecular orbitals ($\Delta$) comes from the inter-atomic hopping within a molecule, typically a $\pi$-type overlap. A typical energy scale for this difference between molecular orbitals in an organic molecule is a few eV (see for example \cite{hinze71}). The inter-site hopping comes from the overlap of molecular orbitals on different molecules, and is exponentially suppresed by distance. These energy scales are on the order of 10 - 100 meV for nearest neighbour overlaps (see for example \cite{jacko13tmttf}).
Thus it is often the case in such systems that $|\Delta| > |t_a| + |t_b|$; the bands resulting from each molecular orbital are narrow enough and well-separated enough that they do not overlap in energy. Thus, depending on filling, one can consider just one orbital or the other as the foundation for an effective model Hamiltonian.

In applying the Wannier construction procedure outlined above, 
 we have glossed over the details of limiting the Fourier transform window to some small energy range. In the simplest case of a single band system, it is clear that this originally-infinite window can be truncated to be exactly the bandwidth of the single band without any loss of generality. However, in multi-band systems, extracting a subset of bands can become difficult. To understand why, let us consider again the 1D, two orbital model.

For $|\Delta| \leq |t_a| + |t_b|$, there are gapless excitations possible between these bands; the two bands have weight in an overlapping energy range. Thus, even in this simplest case without band crossings or interactions, the presence of a second band makes picking out the states of the first band non-trivial. When these bands cross or hybridise, this further complicates the procedure. This problem is very difficult to solve in general and is known as the disentangling problem.

Now, the separation of energy scales in molecular crystals comes in to play. As discussed above, because of the often quite different energy scales of inter- and intra-molecular interactions, it is typical to find well isolated sets of bands in the band structure of a molecular crystal, as illustrated in Fig. \ref{fig:bsdos}. 
This property means that one can bypass the difficulty and ambiguity of projective disentangling procedures. 
Thus minimal input is needed into the WO construction procedure in molecular crystals; one just inputs how many orbitals you would like, spanning what energy window.

\section{\textit{ab initio} Model Construction}

It is important when modeling these systems that the models we use be as accurate and unbiased as possible; starting with preconceptions of how the model `should’ look can limit what one finds.

\subsection{Tight-Binding Models}
DFT can give us useful information about the non-interacting electronic properties of a system. To utilise this information, we will construct a tight-binding model from the DFT using a rigorous Wannier orbital construction technique.
This procedure creates localised orbitals that accurately represent the electronic properties found by DFT for the frontier electrons, those that determine the low-temperature physics.  
As discussed above, the separation of energy scales makes Wannier orbital construction straightforward in molecular crystals. 
Once one has local Wannier orbitals one can construct a first principles tight-binding model.

\subsubsection{First principles versus Fitting}
One might ask why is all of this effort justified? Why not simply write down a perfectly good tight binding model and fit it to a first principles band structure? (As is often done, see \cite{kandpal09,scriven12,jeschke12,seo15} for just a sample.)
For very simple systems, where the relevant tight-binding model is clear, and only has a few parameters, a first principles method is probably not justified as the band structure is well reproduced by fitting methods  \cite{kandpal09,scriven12,jeschke12,seo15}. However, in the much more common situation of a somewhat ambiguous tight-binding model with an unknown number of relevant parameters, a first principles approach is ideal, as I will discuss further below. In fact, in some systems where a quite simple model seemed obvious, it has been shown that a somewhat more nuanced model does a better job of capturing the important many-body physics of the material (as discussed in the context of \pdmit, below).

\begin{figure}
\begin{center}
 \includegraphics[width=0.95\columnwidth]{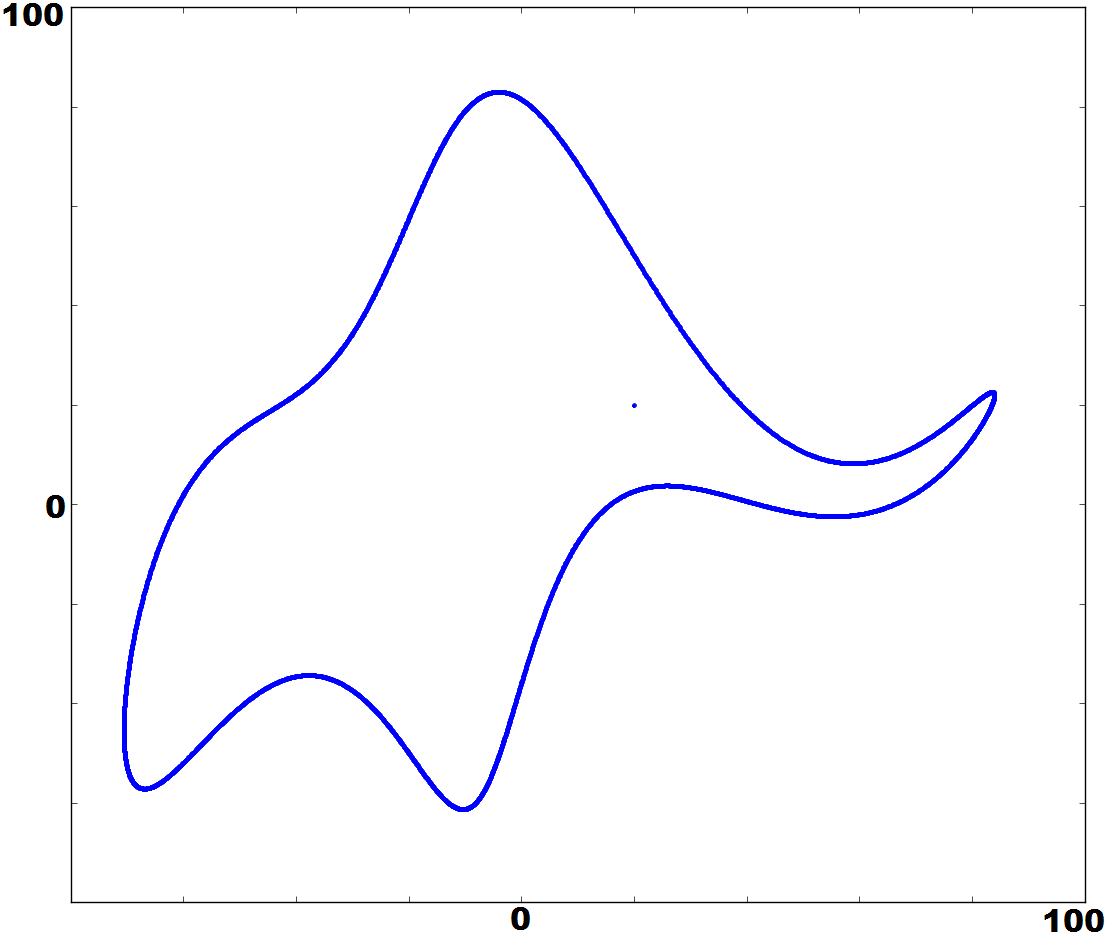} 
 \caption{A four-parameter fit to an elephant, produced as described in \cite{mayer10}. A fifth parameter does indeed allow it to wiggle its trunk. While conforming to the letter of the statement, this implementation somewhat defies its spirit as the parameters are complex numbers, thereby carrying twice the information of four real numbers.}\label{fig:elephant}
\end{center}
\end{figure}

John von Neumann is famously claimed to have said ``With four parameters I can fit an elephant, and with five I can make him wiggle his trunk.'' \cite{dysononfermi} (this is more-or-less true, as shown in Fig. \ref{fig:elephant}, after Ref. \cite{mayer10}); this captures the essence of the problem of fitting energy bands to a tight-binding model. With enough parameters in the fit, it is difficult to have a \emph{bad} fit of the band structure; a model with many parameters might reproduce the dispersion without having any connection to a realistic microscopic description of the system. As such, a good fit provides almost no information, especially not about the quality of the microscopic model to which you are fitting. Worse, when there are many parameters, very different values can produce similarly good fits by whatever optimisation metric you are using. Often, these different sets of values have importantly different physical consequences (for example changing the electronic dimerisation of a chain, or the localisation of charge). These differences can lead to importantly different many-body ground states, as will be discussed further below.

\begin{figure}
\begin{center}
 \includegraphics[width=0.9\columnwidth]{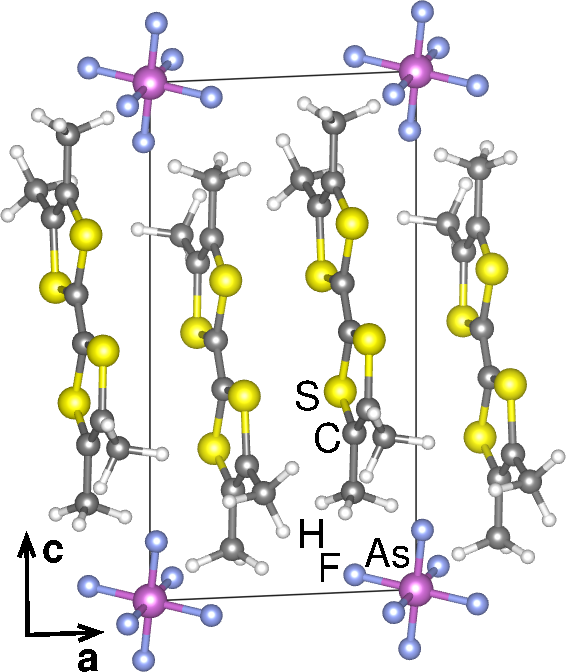} 
 \caption{Crystal structure of TMTTF$_2$AsF$_6$ at room temperature, showing a $\pi$-stacked chain along the $a$ direction, and spacing due to anions in the $c$ direction. Reproduced from \cite{jacko13tmttf}.}\label{fig:tmttfstructure}
\end{center}
\end{figure}

To demonstrate this, I will discuss a particular case of producing a tight-binding model for an organic molecular crystal by fitting and via Wannier orbitals. TMTTF$_2$AsF$_6$ is one of the Fabre salts, a family of organic charge-transfer salts with a rich phase diagram, given in Fig. \ref{fig:fabrephase}. These flat organic molecules $\pi$-stack into one dimensional chains along the crystallographic $a$ direction (shown in Fig. \ref{fig:tmttfstructure}), with some inter-chain coupling in the $a-b$ plane, and minimal coupling in the $c$ direction (where the anionic AsF$_6$ layers introduce a large spacing between TMTTF layers). Thus this system is largely 1D electronically, with some 2D nature introduced by next-nearest neighbour hopping and further terms \cite{nogami05,jacko13tmttf,brown15}. 

In this family of materials, once one decides to include any 2D hopping terms, one encounters the problem that there are many terms of similar magnitude; to avoid neglecting terms of the order one keeps, one needs to add many parameters to the tight-binding model. These many degrees of freedom cause problems for fitting procedures; one finds many local minima with similar optimiziation functions values, but very different parameters, with important physical consequences. This is the heart of the problem; that such fits are examples of \textit{sloppy models}\cite{transtrum15}: changes in one parameter can be almost completely masked by compensatory changes in other parameters.

\begin{figure}
\begin{center}
 \includegraphics[width=0.95\columnwidth]{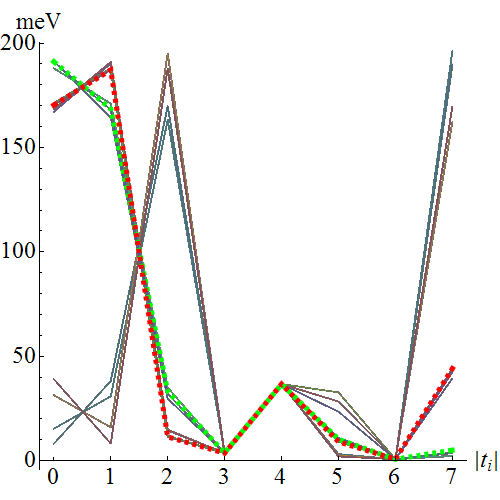} 
 \includegraphics[width=0.95\columnwidth]{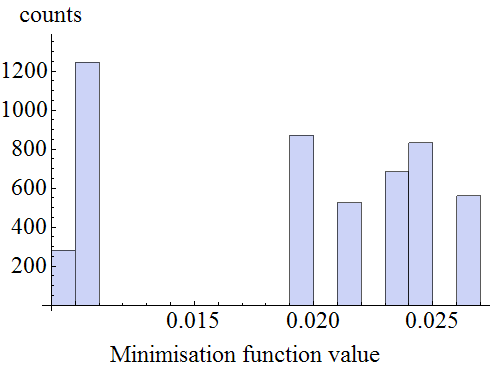} 
 \caption{Instability of fitting with many parameters. 5000 fits of an 8 $t$ tight-binding model to the band structure of TMTTF$_2$AsF$_6$ produced in \cite{jacko13tmttf}. Each line is a set of parameters resulting from one run of the fitting algorithm. The best fit parameters are shown in red. They are inconsistent with the parameters found from Wannier orbital overlaps (given in green). Only 1.5\% of the runs found the minimal value of the objective function. The histogram shows the sets of minimisation function values produced in the 5000 runs. Less than 6\% of runs are in the `best' segment of the histogram, and 1/4 in the best two segments.}\label{fig:tbfitting}
\end{center}
\end{figure}

\begin{figure}
\begin{center}
 \includegraphics[width=0.9\columnwidth]{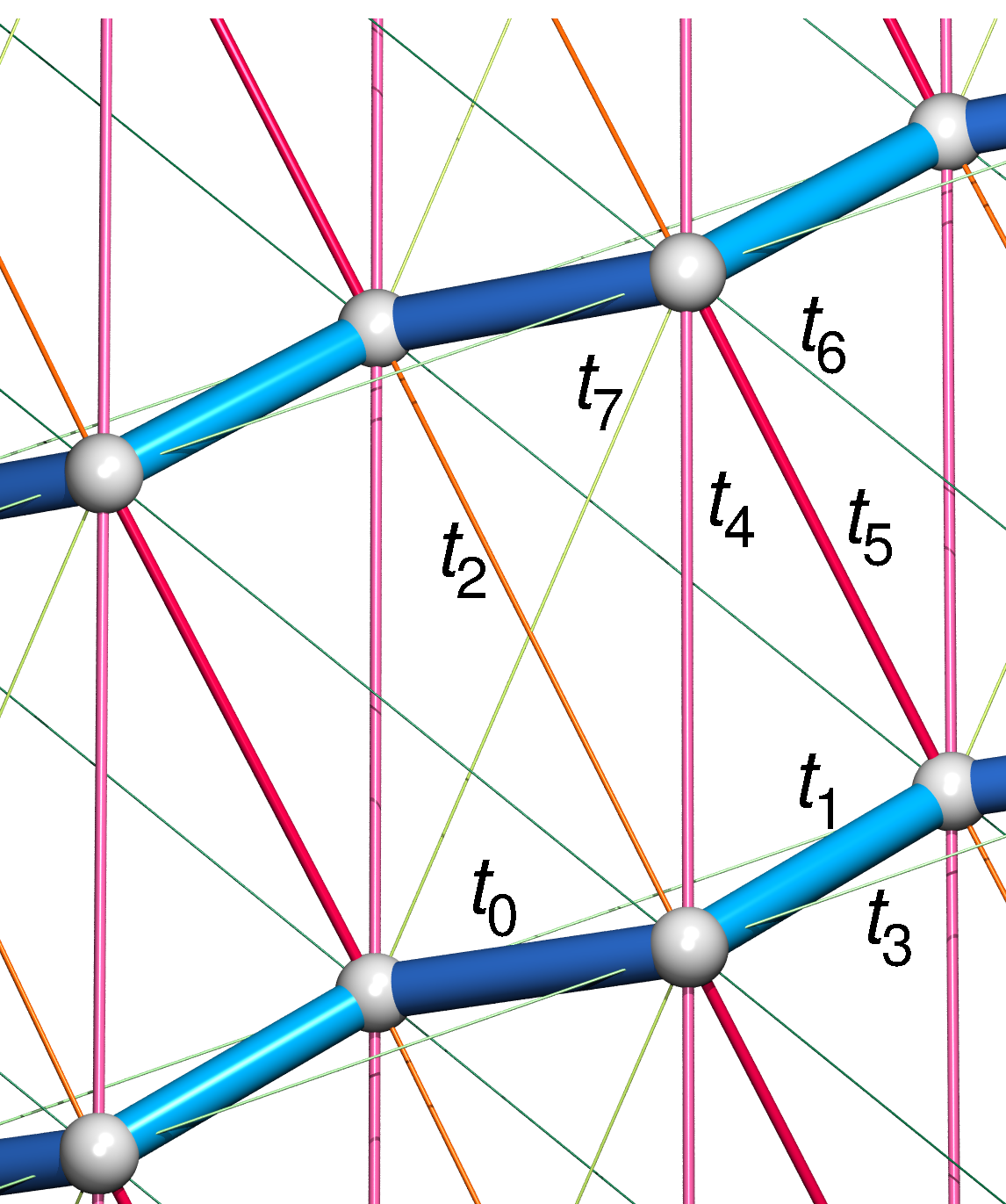} 
 \caption{Tight-binding lattice for TMTTF$_2$AsF$_6$ produced from Wannier orbitals. The thickness of the lines is proportional to the magnitude of the $t$; $t_0 = 175$ meV, $t_1 = 157$ meV, and the rest are $< 25$ meV. Reproduced from \cite{jacko13tmttf}.}\label{fig:tmttfnetwork}
\end{center}
\end{figure}

Figure \ref{fig:tbfitting} shows the results of 5000 runs of a fitting procedure applied to the band structure of TMTTF$_2$AsF$_6$ \cite{jeschkecode, jacko13tmttf}; a pre-defined tight binding lattice is input, and the values of the 8 different hopping integrals (c.f. Fig. \ref{fig:tmttfnetwork}) are optimised to provide the best fit to the band structure (quantified by the least-squares error over the set of points the bands are sampled on). Each run starts with a (different) randomised set of fitting parameters and iterated on.
Due to the nature of this method, each of the 5000 fits is basically unique. Rounding each $t$ to the nearest 0.1 meV, there are 26 unique fits; of the 5000 results, 74 have the minimum objective function. A given run has a 1.5\% chance (1/67) of finding this `best' solution!
There are many more fits with a just slightly higher value of the objective function, and these fits contain contradictory physical information. For example, these different fits make different predictions about the electronic dimerisation of the system; whether the electronic dimers are on the structural dimers or not. This is an important difference, and can change as a function of pressure for example \cite{jacko13tmttf}.
While most of the parameters have positive and negative equivalents, a few parameters are very precisely determined, and with a fixed phase. This means that the relative phases of the hopping integrals cannot be absorbed by a gauge transformation; these different sets of parameters have different physical meanings.
The number of parameters used in the fit also effects the outcome. Removing some $t$ parameters from the inputed model will cause the other $t$ values to change. 
Damningly, in this system the optimal tight-binding fit is quite different to the set of hopping integrals found from Wannier orbitals, as shown in Fig. \ref{fig:tbfitting}. The Wannier and fitted parameters make contraditary predictions about the electronic dimerisation of the system (as seen in the magnitude of the first two $t$'s). In the limit of including all Wannier overlaps out to infinite distance, the set of Wannier tight-binding parameters will exactly reproduce the band structure when the bands are not entagled.

Overall, it is hard to take fitting procedures too seriously with more than a couple of parameters in the fit. With small numbers of parameters, fitting becomes more stable, and in certain systems a few parameters is enough to acurately describe the band structure \cite{kandpal09,jeschke12,seo15}. Even then, a fitted model should be considered an effective model that has potentially lost important detail in `integrating out' the full set of parameters. In a sense these are `variational Hamiltonians'; they are optimised by some metric, but there is no assurance that they represent the underlying microscopic physics.

\begin{figure}
\begin{center}
 \includegraphics[height=0.7\columnwidth, angle=-90]{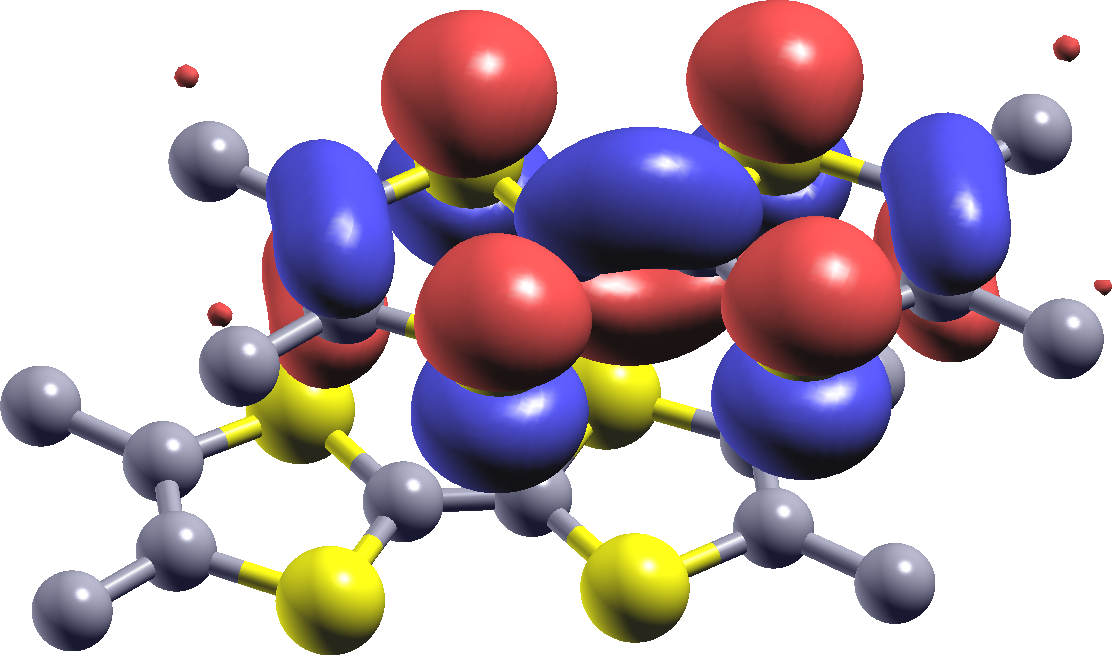} 
 \caption{Wannier orbital for TMTTF$_2$AsF$_6$. Note that this orbital is localised to a single molecules, and very much like the HOMO of an isolated TMTTF molecule. Having this real-space orbital allows us to do many further computations based on the DFT, such as computing a tight-binding model by taking real space overlaps of such orbitals. Reproduced from \cite{jacko13tmttf}.}\label{fig:tmttfWF}
\end{center}
\end{figure}

By producing Wannier orbitals for molecular crystals, and computing a set of $t$'s from those, one finds a single set of parameters that is reliable and robust; one can believe them just as much as one believes the other results of the DFT computation. In the general case of Wannier orbital construction, there is ambiguity involved in disentangling bands to produce the desired set of Wanniers \cite{marzari12}. When the bands one cares about are well-separated from the bulk, this ambiguity is gone. Not only that, but one can gain knowledge by looking at the Wannier orbitals themselves. In the case of TMTTF$_2$AsF$_6$, the Wannier orbitals are localised to single TMTTF molecules, as shown in Fig. \ref{fig:tmttfWF}. 
Here, the Wannier orbital is very much like the HOMO (highest occupied molecular orbital) of a single TMTTF molecule in vacuum. Having this real-space orbital allows us to do many further computations based on the DFT, as well as producing a single robust parameter set for a tight-binding model (Fig. \ref{fig:tmttfnetwork}). This Wannier based model construction technique is being used more and more in molecular systems \cite{nakamura09,nakamura12,jacko13tmttf,jacko13dmit,altmeyer15}.

\subsection{Including Interactions}

Correctly parameterising many-body effects is an important and challenging task; it is the competition between energy scales that leads to the interesting physics in many systems, so small relative changes in large parameters can have large effects \cite{jacko10a}. Often, these parameters are estimated without careful consideration of the assumptions involved. For instance, if one considers a Hubbard model on a dimer with an inter-monomer hopping $t$, on-monomer Hubbard repulsion $U_m$, and inter-monomer Hubbard repulsion $V$; in the limit $U_m \rightarrow \infty$, $V \rightarrow 0$, the effective Hubbard repulsion in the dimer orbitals is $U_d = 2 t$ \cite{komatsu96, mckenzie98}. 
This assumption is often used, since it allows one to estimate many-body parameters from straightforward band structure calculations, or molecular H{\"u}ckel calculations. However, it is not well justified.
It has since been shown that in the more realistic case of $U_m \sim V \gg t$, that $U_d = \frac{1}{2}(U_m + V)$ \cite{scriven09b}. Thus, one still needs to be able to correctly compute many-body parameters to estimate the dimer parameters (even before considering screening). None-the-less, this approximation continues to be used (for example \cite{jeschke12,yoshimi12,tsumuraya13}), often without stating the strong underlying assumptions.

One might think that, given these real-space Wannier orbitals for a particular system, it must be straightforward to calculate the many-body Coulomb integrals and parameterise a Hubbard model. However, computing these terms by simply evaluating the Coloumb energy for each orbital neglects screening (equivalently, relaxation of the bulk states). Screening can easily suppress the Hubbard $U$ by an order of magnitude \cite{nakamura06}.

In classical electromagnetics, there are many techniques for determining the response of a bulk to a perturbing field (analogous to the case here of computing the screening/relaxation of a doubly occupied orbital). The discrete dipole approximations (also called the coupled dipole approximation) is one such technique. In this approximation, one discretises the bulk as a set of polarisable dipoles, and self-consistently solves their response to the perturbing field and to each other \cite{devoe64}. This method is very much like a technique applied to molecular crystals to compute screened Coloumb parameters. By representing each molecule by a set of polarisable (classical) dipoles, and placing a perturbing charge on one lattice site, one can compute the correction to the Hubbard repulsion due to the polarisation of the rest of the molecules in the crystal \cite{cano10b}. This technique, though promising, has only been applied to a single molecular crystal (TTF-TCNQ), with no new applications apparent in the 5 years since the original publication.

An alternative approach to computing screened Coulomb parameters from first principles has gained prominence recently, the constrained random phase approximation (cRPA) \cite{aryasetiawan04}. This technique is also based around computing the polarisation of the system, but in this case, the quantum mechanical polarisation function in the random phase approximation. Here we discuss RPA, cRPA, and its application to molecular crystals in more detail. Practically, it also relies on having Wannier orbitals for a few relevant bands, and so like the tight-binding model construction it is particularly suitable to molecular crystals.

\subsubsection{Constrained Random Phase Approximation}

\begin{figure}
\begin{center}
 \includegraphics[width=0.95\columnwidth]{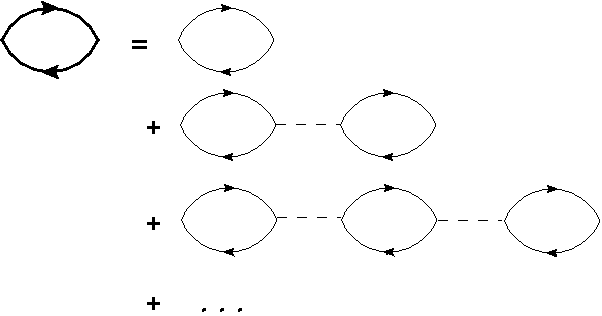} 
 \caption{Random phase approximation bubble diagrams appropriate for calculating the polariation function. Bold lines are fully interacting Greens functions, while non-bold are non-interacting \cite{rpabubblewiki}.}\label{fig:rpabubbles}
\end{center}
\end{figure}

The random phase approximation (RPA) was introduced by David Bohm and David Pines in the 1950's to include the effects of screening into models of electron gases \cite{bohm51,pines52,bohm53,pines53}. Murray Gell-Mann and Keith Brueckner placed this approximation on a firmer footing, showing that the RPA can be derived from a self-consistent series of leading order Feynman diagrams \cite{gellmann57}, an example of which is illustrated in Fig. \ref{fig:rpabubbles}.

Given a basis of occupied and unoccupied states, one can compute an RPA polarisation function
\begin{equation} \label{eq:rpapol}
\begin{split}
&P(\vec{r},\vec{r}',\omega)=\sum^{occ}_{i} \sum^{unocc}_{j} \psi_i(\vec{r}) \psi_i^*(\vec{r}') \psi_j^*(\vec{r}) \psi_j(\vec{r}') \times \\
 & \left( \frac{1}{\omega - \varepsilon_j + \varepsilon_i + i0^{+}} -  \frac{1}{\omega + \varepsilon_j - \varepsilon_i - i0^{+}} \right),
\end{split}
\end{equation}
where $i$ ($j$) runs over the occupied (unoccupied) single particle states.  With this polarisation function one can compute the screening effects of this system. This is exactly what one wants if computing the effects of an impurity in such a system, for example. However, if one wants to include many-body effects in a lattice model, then this constitutes an overcounting of the effect.

In 2004, Aryasetiawan \textit{et al.} introduced a new, precise method for constructing sets of screened effective model parameters for strongly correlated lattice models \cite{aryasetiawan04}. The constrained random phase approximation (constrained RPA or cRPA) is a systematic way of accounting for screening in the many-body parameters computed for some basis orbitals. The system is divided into two fragments; the active subspace (often labelled `d'), the space spanned by the orbitals of interest; and the rest of the bands (labelled `r'). 
On a conceptual level, this procedure computes the effects of transitions involving the `r' subspace with RPA (by computing the polarisation function due to these transitions), while leaving the transitions within the `d' subspace to be dealt with in the many-body model that results \cite{aryasetiawan04}. This proceedure allows one to generate all the terms resulting from the Coulomb interaction, on- and off-site repulsive and magnetic interactions. Practically, one constrains the sums in Eq. \ref{eq:rpapol} to exclude transitions within the active subspace.

The partitioning idea at the core of cRPA works best in the same situation that the Wannier orbital proceedure itself works best: a set of relevant bands well separated from the bulk. In the situation of entangled bands; where the natural basis one would like to use mixes with bands due to other states; one can apply the Wannier disentangling proceedure to construct a disentangled basis \cite{miyake09}. In inorganic system, there are difficulties in disentangling the target bands from the bulk. Nonetheless this approach was quickly applied to transition metal systems and simple transition metal oxides \cite{nakamura06,miyake08,nakamura08, miyake09}. 

This approach has been applied to only a few organic crystals \cite{nakamura09,nakamura12}. In those cases where it has, it finds sometimes importantly different parameter values. In the ET charge transfer salt $\kappa$-(ET)$_2$ Cu$_2$ (CN)$_3$, cRPA predicts a value of $U/t$ for the dimer about twice as large as was estimated from a H\"uckel analysis of a dimer (using an optial conductivity estimate of the monomer value, $U_m$), $U_{d}^{cRPA}/t \sim 15$ vs $U_{d}^{Huckel}/t \sim 7$ \cite{komatsu96, nakamura09}. 
 In a simple Hubbard model, this would place this material well into the insulating phase, contrary to the observed metallic behaviour. The cRPA analysis also showed that off-site $V$ terms are significant, $V/U \sim 0.5$, meaning that to properly understand the system one must consider an extended Hubbard model \cite{nakamura09}. While the optical conductivity estimate for the monomer $U_m$ is quite reliable, the assumptions in using this to estimate $U_d$ are not. This Wannier-based approach gives us a reliable first principles estimate of all the Hamiltonian parameters on the same footing.

\section{First priciples approach finds important differences}

Here we discuss particular examples to show that using a first principles approach can give importantly different results and insights than a standard fitting approach; be it caused by subtleties of parameter variations or qualitatively different lattices.

\subsection{\dmitSL: Fine details matter}
To demonstrate the importance of finding a robust set of model parameters we will turn to the example of {\dmitSL}, a spin-liquid candidate material and part of a family of organic molecular crystals with a rich phase diagram; as well as the spin-liquid phase, these materials have Mott insulating, superconducting, spin density wave and valence bond solid phases \cite{powell11,kobayashi91,seya95,tamura02,itou08}. Constructing a coherent picture of this family of materials and their many phases is highly challenging. This effort has been hindered by the fact that, in the usual development of microscopic models, many approximations are made without fully understanding their consequences \cite{jacko13dmit}.

\begin{figure}
\begin{center}
 \includegraphics[width=0.95\columnwidth]{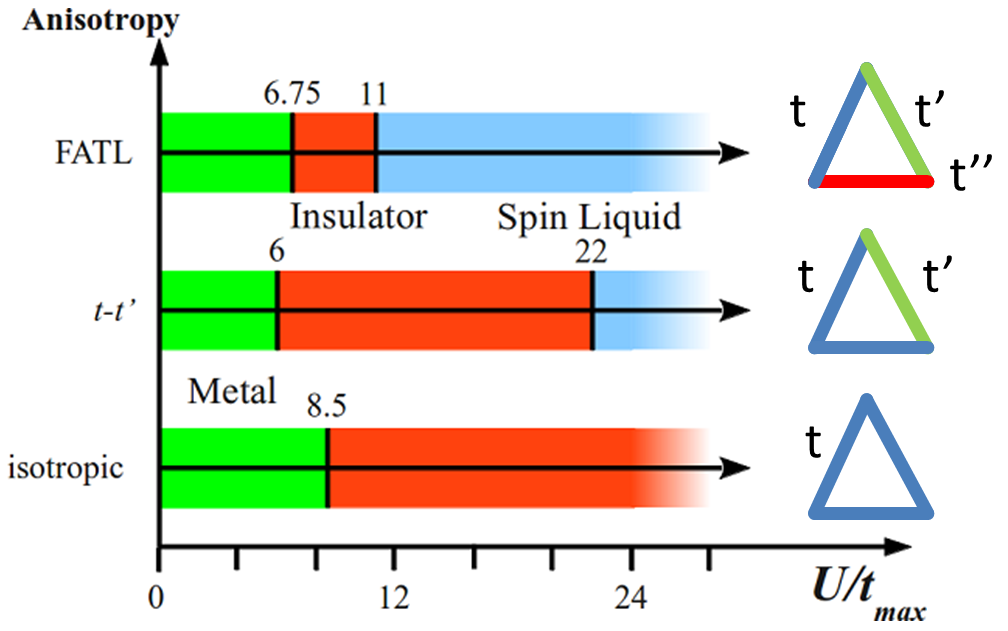} 
 \caption{Phase diagrams of the Hubbard model on the isotropic triangular lattice, $t-t'$ triangular lattice, and fully anisotropic triangular lattice (FATL), for parameters consistent with {\dmitSL}. Note that the transition to the spin-liquid phase occurs for a much smaller value of $U$ in the FATL. Phase diagram determined with variational quantum Monte Carlo \cite{jacko13dmit}. Reproduced from \cite{jacko13dmit}.}\label{fig:fatlphase}
\end{center}
\end{figure}

The typical approach in {\dmitSL} and the related family of materials is to focus on a dimer model, where the dimers of Pd(dmit)$_2$ sit on a $t-t'$ triangular lattice. Parameters are either fit or mapped to this non-interacting model before many-body effects are considered  (see for example Refs. \cite{itou08,scriven12}).
It has since been shown that a fully-anisotropic triangular lattice (FATL; $t-t'-t''$) better represents the electronic structure \cite{tsumuraya13, jacko13dmit,rudra14}. Further, it was shown that a FATL allows one to reproduce the observed many-body properties,  
predicting a spin-liquid ground state for reasonable parameter values in {\dmitSL}, while the $t-t'$ model does not \cite{jacko13dmit}.
Fig. \ref{fig:fatlphase} shows the phase diagram for the Hubbard model (as a function of $U/t$) on the isotropic triangular lattice, $t-t'$ triangular lattice, and fully anisotropic triangular lattice (FATL), each with tight-binding parameters consistent with {\dmitSL} (computed with variational quantum Monte Carlo). First principles estimates predict $U/t_{max} \sim 11$ \cite{nakamura12}. The FATL enters the spin-liquid phase at this point, while the $t-t'$ and isotropic lattices would predict an insulating phase, with this value of $U$ very far from the critical value. Generally, the extra anisotropy seems to destablise the insulating phase relative to the metallic and spin-liquid phases. It is worth noting that these variational quantum Monte Carlo results are not definitive; however, if nothing else, they are indicitive of the important consiquences that even slight parameter changes can have.

Such highly anisotropic models have since become increasingly used in investigations of this family of materials \cite{seo15}. Having reliable and believable predictions of the degrees of anisotropy in these materials (were the fine variation in parameter values can be attributed to physics and not a quirk of the particular fit one is applying) will be vital for building an understanding of the whole class of materials.

\subsection{$\kappa$-(BEDT-TTF)$_2$ salts: Long range terms}
The BEDT-TTF (bis(ethylenedithio)-tetrathiafulvalene, or ET) organic charge transfer salts are a family of quasi-2D crystals that exhibit a wide range of strongly correlated phases (such as non-BCS $d$-wave superconductivity) \cite{mayaffre95,kino96,vorontsov07,powell11,jacko13et,brown15}.
Understanding this wide range of phases requires a good effective model and good model parameters. 
It was in these materials that the shortcomings of the $U \sim t_{intra}$ approximation (discussed above) were made clear, showing that it leads to a systematic underestimate of $U$ \cite{scriven09b}. Once a realistically large value of $U$ is used (computed with cRPA and found to be a 50\% - 100\% increase over previous estimates), a straightforward Hubbard model of the dimer lattice does not capture anything but the Mott insulating phase \cite{nakamura09}. These cRPA parameter estimates also showed that the nearest neighbour inter-site Coulomb interactions are significant ($V/U \sim 0.5$), and that they decay slowly with distance \cite{nakamura09}. 

By applying first principles model building techniques, it was found that describing the phases of the ET salts requires models like the extended Hubbard model with significant and long-ranged inter-site interactions. This kind of model, although it has more parameters, it has no more \textbf{free} parameters. Additionally, the inclusion of long-range Coulomb terms has important implications for the energetics of ordered phases \cite{nakamura09}.

\subsection{{\moly}: An unexpected lattice}
In the previous examples, we showed how details in the model parameters are found to have significant consequences on the predicted ground state. We now turn to a system where, by using a first principles method, one finds a totally different lattice model than any previously considered for this system. 

\begin{figure}
\begin{center}
 \includegraphics[width=0.9\columnwidth]{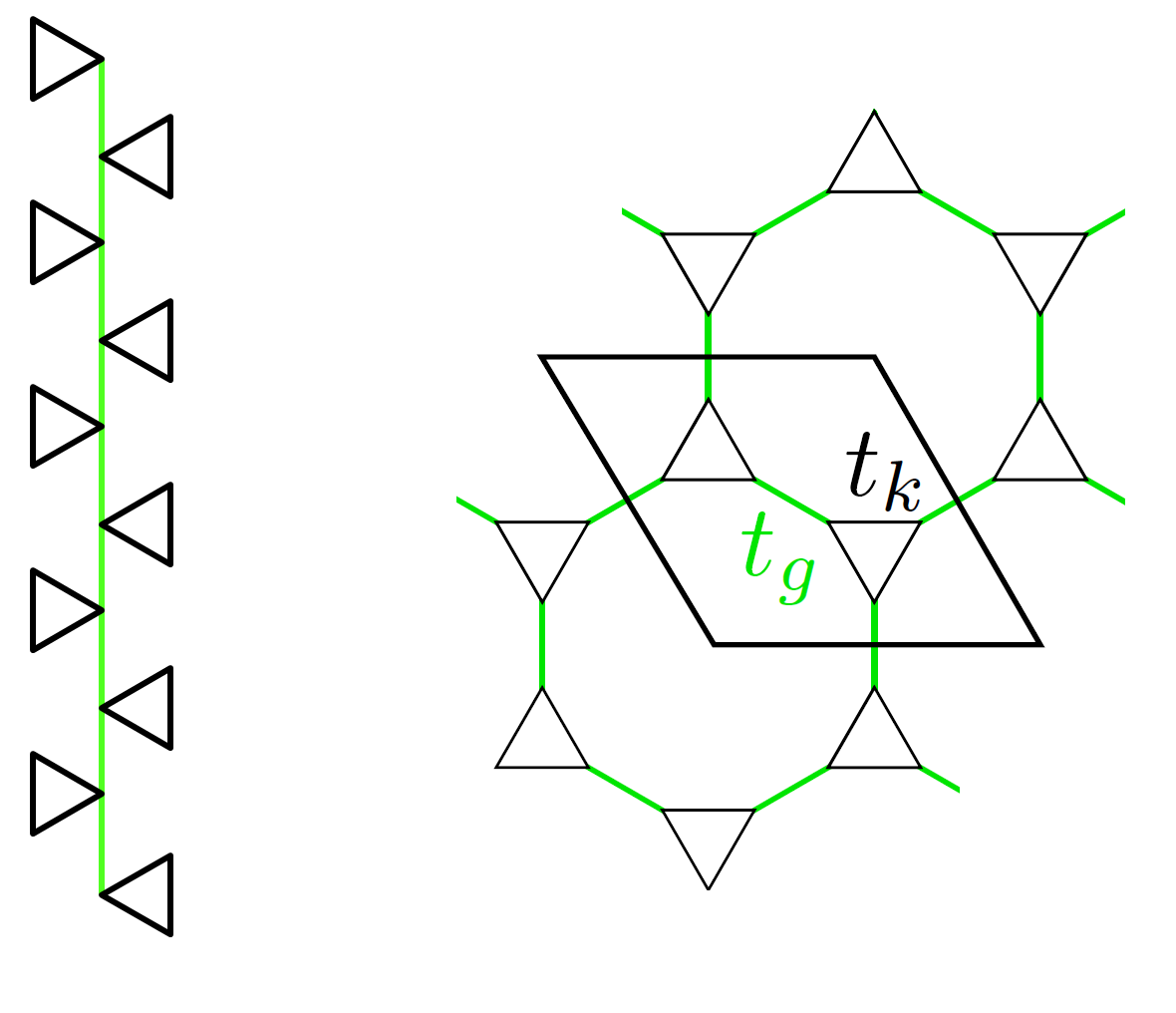} 
 \caption{Lattices for {\moly}: the phenomenological `triangular necklace' lattice on the left, and first priciples kagomene on the right. While quite different, both have interesting topological properties and can provide insights into the behaviour of {\moly}.}\label{fig:mo3s7lattices}
\end{center}
\end{figure}

{\moly} is a single component molecular crystal that was designed to be metallic. However, it was found to be an activated insulator with an activation energy of 34 meV \cite{llusar04}. Further, it was found to have no sign of any magnetic order down to very low temperatures ($J/k_B T \sim 50$ \cite{llusar04}); a possible sign of a spin-liquid state. Based on the apparent 1D physicical properties, its crystal structure, and initial bandstructure calculations, {\moly} was modeled with a one-dimensional lattice \cite{llusar04,janani14a,janani14b}. 
This is the 1D `triangular necklace' lattice, illustrated in Fig. \ref{fig:mo3s7lattices}. The ground state of the Hubbard model on this lattice at $2/3$ filling (as appropriate for {\moly}) is found to be in the Haldane phase, consistant with the experimental evidence \cite{janani14a, janani14b}.

However, Wannier orbital tight-binding model construction based on density functional theory for {\moly} predicts that at the single electron level, this system is actually 2D with coupling between the 2D layers. The lattice of the 2D layers is an unusual decorated honeycomb lattice, the `kagomene' lattice (illustrated in Fig. \ref{fig:mo3s7lattices}); interpolating between the graphene (honeycomb) and kagom{\'e} lattices \cite{jacko15a}. These lattices have quite different properties, and provide quite different pictures of the physics of {\moly}. The kagomene lattice has been studied theoretically before \cite{yao07,wen10,ruegg10,tikhonov10,yao13}, but never seen in a real system. This first priciples approach found a layered kagomene lattice in {\moly} quite unexpectedly, demonstrating the novel insights this appoach can yield. The microscopic picture produced is quite different from the phenomenological model.

\begin{figure}
\begin{center}
 \includegraphics[width=0.9\columnwidth]{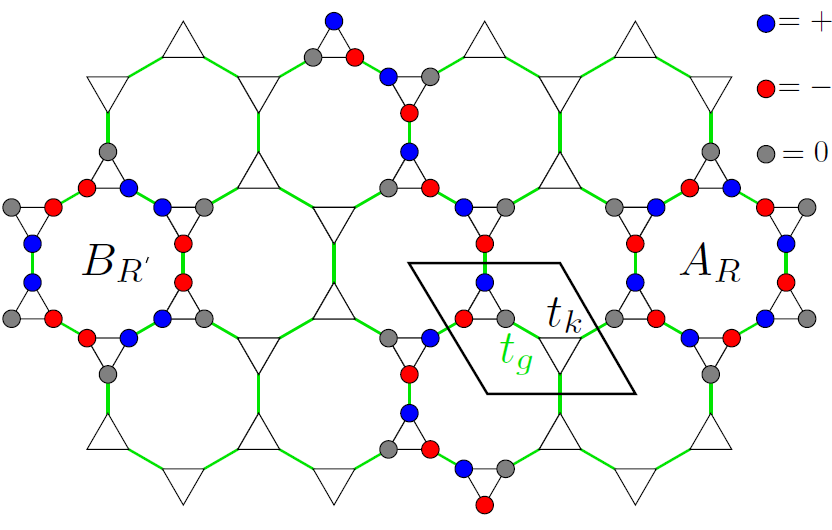} 
 \caption{Localised states on the kagomene lattice, showing two plaquette states, the anti-bonding $A_{\bf R}$ and bonding $B_{{\bf R}'}$, and a topologically non-trivial loop state, which can only exist with periodic boundary conditions. Reproduced from \cite{jacko15a}.}\label{fig:kagomeneflatband}
\end{center}
\end{figure}

The one dimensional behaviour can be understood on the grounds of the kagomene lattice: just like kagom{\'e}, this lattice has exactly localised states \cite{bergman08}, illustrated in Fig. \ref{fig:kagomeneflatband}. Once the 2D kagomene lattice is extended into 3D, these localised states become 1D bands. These emergent 1D states are topological; their degeneracy depends on the boundary conditions of the lattice. Thus, dispite the hopping integrals having similar magnitudes in every direction (in fact, slightly smaller in the stacking direction), one recovers the 1D behaviour and gains some important insights about the potential topological properties of this system.

As a phenomenological model, the necklace lattice does a good job of reproducing the observed magnetic properties of {\moly} \cite{janani14b}. On the other hand, the kagomene model provides a natural explanation for the quasi-1D behaviour, and highlights the interesting topological flat bands analogous to those seen in the kagom{\'e} lattice \cite{jacko15a}. In addition, one can find a lattice closely related to the necklace model as a limiting behaviour of an interacting model in the kagomene lattice, and the many-body behaviour of this model is very similar to the necklace model \cite{nourse}. One can naturally find new terms to extend the necklace model in a consistent way by introducing higher-order terms in these limits, for example including the chiral next-nearest neighbour terms \cite{jacko15a}.

\section{Summary}
Over the last decade Wannier orbitals have become an important tool for predictive physics. By constructing Wannier orbitals for frontier bands, we can derive effective models that avoid our biases. These models are robust and reliable, and allow us to make detailed comparisons between materials, and start to extract some general behaviours. They can also find models that we might never have expected to see, leading us to new insights.
By moving to this kind of assumption-free methodology for model construction, we can move to a truly \textbf{pre}dictive approach.
We can avoid the dangers of relying on variational Hamiltonians, and allow ourselves to find truely unexpected things.

\section{Acknowledgments}
I would like to thank Ben Powell, Ross McKenzie, Roser Valent{\'i}, Harald Jeschke, and Klaus Koepernik for many interesting discussions on these topics.
I thank Ben Powell and Ross McKenzie for their helpful comments on this manuscript.
I was supported by the Australian Research Council (ARC) through Grant No. DP130100757.


\begin{thebibliography}{78}
\expandafter\ifx\csname natexlab\endcsname\relax\def\natexlab#1{#1}\fi
\expandafter\ifx\csname bibnamefont\endcsname\relax
  \def\bibnamefont#1{#1}\fi
\expandafter\ifx\csname bibfnamefont\endcsname\relax
  \def\bibfnamefont#1{#1}\fi
\expandafter\ifx\csname citenamefont\endcsname\relax
  \def\citenamefont#1{#1}\fi
\expandafter\ifx\csname url\endcsname\relax
  \def\url#1{\texttt{#1}}\fi
\expandafter\ifx\csname urlprefix\endcsname\relax\def\urlprefix{URL }\fi
\providecommand{\bibinfo}[2]{#2}
\providecommand{\eprint}[2][]{\url{#2}}

\bibitem[{\citenamefont{Chaikin et~al.}(1998)\citenamefont{Chaikin,
  Chashechkina, Lee, and Naughton}}]{chaikin98}
\bibinfo{author}{\bibfnamefont{P.~M.} \bibnamefont{Chaikin}},
  \bibinfo{author}{\bibfnamefont{E.~I.} \bibnamefont{Chashechkina}},
  \bibinfo{author}{\bibfnamefont{I.~J.} \bibnamefont{Lee}}, \bibnamefont{and}
  \bibinfo{author}{\bibfnamefont{M.~J.} \bibnamefont{Naughton}},
  \bibinfo{journal}{Journal of Physics: Condensed Matter}
  \textbf{\bibinfo{volume}{10}}, \bibinfo{pages}{11301} (\bibinfo{year}{1998}).

\bibitem[{\citenamefont{Powell and McKenzie}(2011)}]{powell11}
\bibinfo{author}{\bibfnamefont{B.~J.} \bibnamefont{Powell}} \bibnamefont{and}
  \bibinfo{author}{\bibfnamefont{R.~H.} \bibnamefont{McKenzie}},
  \bibinfo{journal}{Rep. Prog. Phys.} \textbf{\bibinfo{volume}{74}},
  \bibinfo{pages}{056501} (\bibinfo{year}{2011}).

\bibitem[{\citenamefont{Jacko et~al.}(2013{\natexlab{a}})\citenamefont{Jacko,
  Tocchio, Valent{\'\i}, and Jeschke}}]{jacko13dmit}
\bibinfo{author}{\bibfnamefont{A.~C.} \bibnamefont{Jacko}},
  \bibinfo{author}{\bibfnamefont{L.~F.} \bibnamefont{Tocchio}},
  \bibinfo{author}{\bibfnamefont{R.}~\bibnamefont{Valent{\'\i}}},
  \bibnamefont{and} \bibinfo{author}{\bibfnamefont{H.~O.}
  \bibnamefont{Jeschke}}, \bibinfo{journal}{Phys. Rev. B}
  \textbf{\bibinfo{volume}{88}}, \bibinfo{pages}{155139}
  (\bibinfo{year}{2013}{\natexlab{a}}).

\bibitem[{\citenamefont{Dressel}(2007)}]{dressel07}
\bibinfo{author}{\bibfnamefont{M.}~\bibnamefont{Dressel}},
  \bibinfo{journal}{Naturwissenschaften} \textbf{\bibinfo{volume}{94}},
  \bibinfo{pages}{527} (\bibinfo{year}{2007}).

\bibitem[{\citenamefont{Seo et~al.}(2004)\citenamefont{Seo, Hotta, and
  Fukuyama}}]{seo04}
\bibinfo{author}{\bibfnamefont{H.}~\bibnamefont{Seo}},
  \bibinfo{author}{\bibfnamefont{C.}~\bibnamefont{Hotta}}, \bibnamefont{and}
  \bibinfo{author}{\bibfnamefont{H.}~\bibnamefont{Fukuyama}},
  \bibinfo{journal}{Chemical Reviews} \textbf{\bibinfo{volume}{104}},
  \bibinfo{pages}{5005} (\bibinfo{year}{2004}).

\bibitem[{\citenamefont{Brown}(2015)}]{brown15}
\bibinfo{author}{\bibfnamefont{S.}~\bibnamefont{Brown}},
  \bibinfo{journal}{Physica C: Superconductivity and its Applications}
  \textbf{\bibinfo{volume}{514}}, \bibinfo{pages}{279 } (\bibinfo{year}{2015}).

\bibitem[{\citenamefont{Jacko et~al.}(2013{\natexlab{b}})\citenamefont{Jacko,
  Feldner, Rose, Lissner, Dressel, Valent{\'\i}, and Jeschke}}]{jacko13tmttf}
\bibinfo{author}{\bibfnamefont{A.~C.} \bibnamefont{Jacko}},
  \bibinfo{author}{\bibfnamefont{H.}~\bibnamefont{Feldner}},
  \bibinfo{author}{\bibfnamefont{E.}~\bibnamefont{Rose}},
  \bibinfo{author}{\bibfnamefont{F.}~\bibnamefont{Lissner}},
  \bibinfo{author}{\bibfnamefont{M.}~\bibnamefont{Dressel}},
  \bibinfo{author}{\bibfnamefont{R.}~\bibnamefont{Valent{\'\i}}},
  \bibnamefont{and} \bibinfo{author}{\bibfnamefont{H.~O.}
  \bibnamefont{Jeschke}}, \bibinfo{journal}{Phys. Rev. B}
  \textbf{\bibinfo{volume}{87}}, \bibinfo{pages}{155139}
  (\bibinfo{year}{2013}{\natexlab{b}}).

\bibitem[{\citenamefont{Seo et~al.}(2015)\citenamefont{Seo, Tsumuraya,
  Tsuchiizu, Miyazaki, and Kato}}]{seo15}
\bibinfo{author}{\bibfnamefont{H.}~\bibnamefont{Seo}},
  \bibinfo{author}{\bibfnamefont{T.}~\bibnamefont{Tsumuraya}},
  \bibinfo{author}{\bibfnamefont{M.}~\bibnamefont{Tsuchiizu}},
  \bibinfo{author}{\bibfnamefont{T.}~\bibnamefont{Miyazaki}}, \bibnamefont{and}
  \bibinfo{author}{\bibfnamefont{R.}~\bibnamefont{Kato}}, \bibinfo{journal}{J.
  Phys. Soc. Jap.} \textbf{\bibinfo{volume}{84}}, \bibinfo{pages}{044716}
  (\bibinfo{year}{2015}).

\bibitem[{\citenamefont{Coldea et~al.}(2010)\citenamefont{Coldea, Tennant,
  Wheeler, Wawrzynska, Prabhakaran, Telling, Habicht, Smeibidl, and
  Kiefer}}]{coldea10}
\bibinfo{author}{\bibfnamefont{R.}~\bibnamefont{Coldea}},
  \bibinfo{author}{\bibfnamefont{D.~A.} \bibnamefont{Tennant}},
  \bibinfo{author}{\bibfnamefont{E.~M.} \bibnamefont{Wheeler}},
  \bibinfo{author}{\bibfnamefont{E.}~\bibnamefont{Wawrzynska}},
  \bibinfo{author}{\bibfnamefont{D.}~\bibnamefont{Prabhakaran}},
  \bibinfo{author}{\bibfnamefont{M.}~\bibnamefont{Telling}},
  \bibinfo{author}{\bibfnamefont{K.}~\bibnamefont{Habicht}},
  \bibinfo{author}{\bibfnamefont{P.}~\bibnamefont{Smeibidl}}, \bibnamefont{and}
  \bibinfo{author}{\bibfnamefont{K.}~\bibnamefont{Kiefer}},
  \bibinfo{journal}{Science} \textbf{\bibinfo{volume}{327}},
  \bibinfo{pages}{177} (\bibinfo{year}{2010}).

\bibitem[{\citenamefont{Taniguchi et~al.}(1999)\citenamefont{Taniguchi,
  Kawamoto, and Kanoda}}]{taniguchi99}
\bibinfo{author}{\bibfnamefont{H.}~\bibnamefont{Taniguchi}},
  \bibinfo{author}{\bibfnamefont{A.}~\bibnamefont{Kawamoto}}, \bibnamefont{and}
  \bibinfo{author}{\bibfnamefont{K.}~\bibnamefont{Kanoda}},
  \bibinfo{journal}{Phys. Rev. B} \textbf{\bibinfo{volume}{59}},
  \bibinfo{pages}{8424} (\bibinfo{year}{1999}).

\bibitem[{\citenamefont{J\'erome}(1991)}]{jerome91}
\bibinfo{author}{\bibfnamefont{D.}~\bibnamefont{J\'erome}},
  \bibinfo{journal}{Science} \textbf{\bibinfo{volume}{252}},
  \bibinfo{pages}{1509} (\bibinfo{year}{1991}).

\bibitem[{\citenamefont{Marzari et~al.}(2012)\citenamefont{Marzari, Mostofi,
  Yates, Souza, and Vanderbilt}}]{marzari12}
\bibinfo{author}{\bibfnamefont{N.}~\bibnamefont{Marzari}},
  \bibinfo{author}{\bibfnamefont{A.~A.} \bibnamefont{Mostofi}},
  \bibinfo{author}{\bibfnamefont{J.~R.} \bibnamefont{Yates}},
  \bibinfo{author}{\bibfnamefont{I.}~\bibnamefont{Souza}}, \bibnamefont{and}
  \bibinfo{author}{\bibfnamefont{D.}~\bibnamefont{Vanderbilt}},
  \bibinfo{journal}{Rev. Mod. Phys.} \textbf{\bibinfo{volume}{84}},
  \bibinfo{pages}{1419} (\bibinfo{year}{2012}).

\bibitem[{\citenamefont{Wannier}(1937)}]{wannier37}
\bibinfo{author}{\bibfnamefont{G.~H.} \bibnamefont{Wannier}},
  \bibinfo{journal}{Phys. Rev.} \textbf{\bibinfo{volume}{52}},
  \bibinfo{pages}{191} (\bibinfo{year}{1937}).

\bibitem[{\citenamefont{Koster}(1953)}]{koster53}
\bibinfo{author}{\bibfnamefont{G.~F.} \bibnamefont{Koster}},
  \bibinfo{journal}{Phys. Rev.} \textbf{\bibinfo{volume}{89}},
  \bibinfo{pages}{67} (\bibinfo{year}{1953}).

\bibitem[{\citenamefont{Kohn}(1959)}]{kohn59}
\bibinfo{author}{\bibfnamefont{W.}~\bibnamefont{Kohn}}, \bibinfo{journal}{Phys.
  Rev.} \textbf{\bibinfo{volume}{115}}, \bibinfo{pages}{809}
  (\bibinfo{year}{1959}).

\bibitem[{\citenamefont{H{\"u}ckel}(1931)}]{huckel31}
\bibinfo{author}{\bibfnamefont{E.}~\bibnamefont{H{\"u}ckel}},
  \bibinfo{journal}{Zeit. f{\"u}r Physik} \textbf{\bibinfo{volume}{70}},
  \bibinfo{pages}{204} (\bibinfo{year}{1931}).

\bibitem[{\citenamefont{Slater and Koster}(1954)}]{slater54}
\bibinfo{author}{\bibfnamefont{J.~C.} \bibnamefont{Slater}} \bibnamefont{and}
  \bibinfo{author}{\bibfnamefont{G.~F.} \bibnamefont{Koster}},
  \bibinfo{journal}{Phys. Rev.} \textbf{\bibinfo{volume}{94}},
  \bibinfo{pages}{1498} (\bibinfo{year}{1954}).

\bibitem[{\citenamefont{Kohn}(1993)}]{kohn93}
\bibinfo{author}{\bibfnamefont{W.}~\bibnamefont{Kohn}}, \bibinfo{journal}{Chem.
  Phys. Lett.} \textbf{\bibinfo{volume}{208}}, \bibinfo{pages}{167}
  (\bibinfo{year}{1993}).

\bibitem[{\citenamefont{Kohn}(1995)}]{kohn95}
\bibinfo{author}{\bibfnamefont{W.}~\bibnamefont{Kohn}}, \bibinfo{journal}{Int.
  J. Quant. Chem.} \textbf{\bibinfo{volume}{56}}, \bibinfo{pages}{229}
  (\bibinfo{year}{1995}).

\bibitem[{\citenamefont{Cloizeaux}(1964)}]{descloizeaux64}
\bibinfo{author}{\bibfnamefont{J.~D.} \bibnamefont{Cloizeaux}},
  \bibinfo{journal}{Phys. Rev.} \textbf{\bibinfo{volume}{135}},
  \bibinfo{pages}{A685} (\bibinfo{year}{1964}).

\bibitem[{\citenamefont{Marzari and Vanderbilt}(1997)}]{marzari97}
\bibinfo{author}{\bibfnamefont{N.}~\bibnamefont{Marzari}} \bibnamefont{and}
  \bibinfo{author}{\bibfnamefont{D.}~\bibnamefont{Vanderbilt}},
  \bibinfo{journal}{Phys. Rev. B} \textbf{\bibinfo{volume}{56}},
  \bibinfo{pages}{12847} (\bibinfo{year}{1997}).

\bibitem[{\citenamefont{Souza et~al.}(2001)\citenamefont{Souza, Marzari, and
  Vanderbilt}}]{souza01}
\bibinfo{author}{\bibfnamefont{I.}~\bibnamefont{Souza}},
  \bibinfo{author}{\bibfnamefont{N.}~\bibnamefont{Marzari}}, \bibnamefont{and}
  \bibinfo{author}{\bibfnamefont{D.}~\bibnamefont{Vanderbilt}},
  \bibinfo{journal}{Phys. Rev. B} \textbf{\bibinfo{volume}{65}},
  \bibinfo{pages}{035109} (\bibinfo{year}{2001}).

\bibitem[{\citenamefont{Mostofi et~al.}(2008)\citenamefont{Mostofi, Yates, Lee,
  Souza, Vanderbilt, and Marzari}}]{mostofi08}
\bibinfo{author}{\bibfnamefont{A.~A.} \bibnamefont{Mostofi}},
  \bibinfo{author}{\bibfnamefont{J.~R.} \bibnamefont{Yates}},
  \bibinfo{author}{\bibfnamefont{Y.-S.} \bibnamefont{Lee}},
  \bibinfo{author}{\bibfnamefont{I.}~\bibnamefont{Souza}},
  \bibinfo{author}{\bibfnamefont{D.}~\bibnamefont{Vanderbilt}},
  \bibnamefont{and} \bibinfo{author}{\bibfnamefont{N.}~\bibnamefont{Marzari}},
  \bibinfo{journal}{Comp. Phys. Comm.} \textbf{\bibinfo{volume}{178}},
  \bibinfo{pages}{685} (\bibinfo{year}{2008}).

\bibitem[{\citenamefont{Koepernik and Eschrig}(1999)}]{koepernik99}
\bibinfo{author}{\bibfnamefont{K.}~\bibnamefont{Koepernik}} \bibnamefont{and}
  \bibinfo{author}{\bibfnamefont{H.}~\bibnamefont{Eschrig}},
  \bibinfo{journal}{Phys. Rev. B} \textbf{\bibinfo{volume}{59}},
  \bibinfo{pages}{1743} (\bibinfo{year}{1999}).

\bibitem[{\citenamefont{Kune\v{s} et~al.}(2010)\citenamefont{Kune\v{s}, Arita,
  Wissgott, Toschi, Ikeda, and Held}}]{kunes10}
\bibinfo{author}{\bibfnamefont{J.}~\bibnamefont{Kune\v{s}}},
  \bibinfo{author}{\bibfnamefont{R.}~\bibnamefont{Arita}},
  \bibinfo{author}{\bibfnamefont{P.}~\bibnamefont{Wissgott}},
  \bibinfo{author}{\bibfnamefont{A.}~\bibnamefont{Toschi}},
  \bibinfo{author}{\bibfnamefont{H.}~\bibnamefont{Ikeda}}, \bibnamefont{and}
  \bibinfo{author}{\bibfnamefont{K.}~\bibnamefont{Held}},
  \bibinfo{journal}{Comp. Phys. Comm.} \textbf{\bibinfo{volume}{181}},
  \bibinfo{pages}{1888} (\bibinfo{year}{2010}).

\bibitem[{\citenamefont{Giannozzi et~al.}(2009)\citenamefont{Giannozzi, Baroni,
  Bonini, Calandra, Car, Cavazzoni, Ceresoli, Chiarotti, Cococcioni, Dabo
  et~al.}}]{giannozzi09}
\bibinfo{author}{\bibfnamefont{P.}~\bibnamefont{Giannozzi}},
  \bibinfo{author}{\bibfnamefont{S.}~\bibnamefont{Baroni}},
  \bibinfo{author}{\bibfnamefont{N.}~\bibnamefont{Bonini}},
  \bibinfo{author}{\bibfnamefont{M.}~\bibnamefont{Calandra}},
  \bibinfo{author}{\bibfnamefont{R.}~\bibnamefont{Car}},
  \bibinfo{author}{\bibfnamefont{C.}~\bibnamefont{Cavazzoni}},
  \bibinfo{author}{\bibfnamefont{D.}~\bibnamefont{Ceresoli}},
  \bibinfo{author}{\bibfnamefont{G.~L.} \bibnamefont{Chiarotti}},
  \bibinfo{author}{\bibfnamefont{M.}~\bibnamefont{Cococcioni}},
  \bibinfo{author}{\bibfnamefont{I.}~\bibnamefont{Dabo}}, \bibnamefont{et~al.},
  \bibinfo{journal}{J. Phys.: Cond. Matt.} \textbf{\bibinfo{volume}{21}},
  \bibinfo{pages}{395502} (\bibinfo{year}{2009}).

\bibitem[{\citenamefont{Gonze et~al.}(2009)\citenamefont{Gonze, Amadon,
  Anglade, Beuken, Bottin, Boulanger, Bruneval, Caliste, Caracas, Côté
  et~al.}}]{gonze09}
\bibinfo{author}{\bibfnamefont{X.}~\bibnamefont{Gonze}},
  \bibinfo{author}{\bibfnamefont{B.}~\bibnamefont{Amadon}},
  \bibinfo{author}{\bibfnamefont{P.-M.} \bibnamefont{Anglade}},
  \bibinfo{author}{\bibfnamefont{J.-M.} \bibnamefont{Beuken}},
  \bibinfo{author}{\bibfnamefont{F.}~\bibnamefont{Bottin}},
  \bibinfo{author}{\bibfnamefont{P.}~\bibnamefont{Boulanger}},
  \bibinfo{author}{\bibfnamefont{F.}~\bibnamefont{Bruneval}},
  \bibinfo{author}{\bibfnamefont{D.}~\bibnamefont{Caliste}},
  \bibinfo{author}{\bibfnamefont{R.}~\bibnamefont{Caracas}},
  \bibinfo{author}{\bibfnamefont{M.}~\bibnamefont{Côté}},
  \bibnamefont{et~al.}, \bibinfo{journal}{Comp. Phys. Comm.}
  \textbf{\bibinfo{volume}{180}}, \bibinfo{pages}{2582} (\bibinfo{year}{2009}).

\bibitem[{\citenamefont{Freimuth et~al.}(2008)\citenamefont{Freimuth,
  Mokrousov, Wortmann, Heinze, and Bl\"ugel}}]{freimuth08}
\bibinfo{author}{\bibfnamefont{F.}~\bibnamefont{Freimuth}},
  \bibinfo{author}{\bibfnamefont{Y.}~\bibnamefont{Mokrousov}},
  \bibinfo{author}{\bibfnamefont{D.}~\bibnamefont{Wortmann}},
  \bibinfo{author}{\bibfnamefont{S.}~\bibnamefont{Heinze}}, \bibnamefont{and}
  \bibinfo{author}{\bibfnamefont{S.}~\bibnamefont{Bl\"ugel}},
  \bibinfo{journal}{Phys. Rev. B} \textbf{\bibinfo{volume}{78}},
  \bibinfo{pages}{035120} (\bibinfo{year}{2008}).

\bibitem[{\citenamefont{Hinze and Beveridge}(1971)}]{hinze71}
\bibinfo{author}{\bibfnamefont{J.}~\bibnamefont{Hinze}} \bibnamefont{and}
  \bibinfo{author}{\bibfnamefont{D.~L.} \bibnamefont{Beveridge}},
  \bibinfo{journal}{Journal of the American Chemical Society}
  \textbf{\bibinfo{volume}{93}}, \bibinfo{pages}{3107} (\bibinfo{year}{1971}).

\bibitem[{\citenamefont{Kandpal et~al.}(2009)\citenamefont{Kandpal, Opahle,
  Zhang, Jeschke, and Valent\'{i}}}]{kandpal09}
\bibinfo{author}{\bibfnamefont{H.~C.} \bibnamefont{Kandpal}},
  \bibinfo{author}{\bibfnamefont{I.}~\bibnamefont{Opahle}},
  \bibinfo{author}{\bibfnamefont{Y.-Z.} \bibnamefont{Zhang}},
  \bibinfo{author}{\bibfnamefont{H.~O.} \bibnamefont{Jeschke}},
  \bibnamefont{and}
  \bibinfo{author}{\bibfnamefont{R.}~\bibnamefont{Valent\'{i}}},
  \bibinfo{journal}{Phys. Rev. Lett.} \textbf{\bibinfo{volume}{103}},
  \bibinfo{pages}{067004} (\bibinfo{year}{2009}).

\bibitem[{\citenamefont{Scriven and Powell}(2012)}]{scriven12}
\bibinfo{author}{\bibfnamefont{E.~P.} \bibnamefont{Scriven}} \bibnamefont{and}
  \bibinfo{author}{\bibfnamefont{B.~J.} \bibnamefont{Powell}},
  \bibinfo{journal}{Phys. Rev. Lett.} \textbf{\bibinfo{volume}{109}},
  \bibinfo{pages}{097206} (\bibinfo{year}{2012}).

\bibitem[{\citenamefont{Jeschke et~al.}(2012)\citenamefont{Jeschke, de~Souza,
  Valent\'{i}, Manna, Lang, and Schlueter}}]{jeschke12}
\bibinfo{author}{\bibfnamefont{H.~O.} \bibnamefont{Jeschke}},
  \bibinfo{author}{\bibfnamefont{M.}~\bibnamefont{de~Souza}},
  \bibinfo{author}{\bibfnamefont{R.}~\bibnamefont{Valent\'{i}}},
  \bibinfo{author}{\bibfnamefont{R.~S.} \bibnamefont{Manna}},
  \bibinfo{author}{\bibfnamefont{M.}~\bibnamefont{Lang}}, \bibnamefont{and}
  \bibinfo{author}{\bibfnamefont{J.~A.} \bibnamefont{Schlueter}},
  \bibinfo{journal}{Phys. Rev. B} \textbf{\bibinfo{volume}{85}},
  \bibinfo{pages}{035125} (\bibinfo{year}{2012}).

\bibitem[{\citenamefont{Dyson}(2004)}]{dysononfermi}
\bibinfo{author}{\bibfnamefont{F.}~\bibnamefont{Dyson}},
  \bibinfo{journal}{Nature (London)} \textbf{\bibinfo{volume}{427}},
  \bibinfo{pages}{297} (\bibinfo{year}{2004}).

\bibitem[{\citenamefont{Mayer et~al.}(2010)\citenamefont{Mayer, Khairy, and
  Howard}}]{mayer10}
\bibinfo{author}{\bibfnamefont{J.}~\bibnamefont{Mayer}},
  \bibinfo{author}{\bibfnamefont{K.}~\bibnamefont{Khairy}}, \bibnamefont{and}
  \bibinfo{author}{\bibfnamefont{J.}~\bibnamefont{Howard}},
  \bibinfo{journal}{American Journal of Physics} \textbf{\bibinfo{volume}{78}},
  \bibinfo{pages}{648} (\bibinfo{year}{2010}).

\bibitem[{\citenamefont{Nogami et~al.}(2005)\citenamefont{Nogami, Ito,
  Yamamoto, Irie, Horita, Kambe, Nagao, Oshima, Ikeda, and
  Nakamura}}]{nogami05}
\bibinfo{author}{\bibfnamefont{Y.}~\bibnamefont{Nogami}},
  \bibinfo{author}{\bibfnamefont{T.}~\bibnamefont{Ito}},
  \bibinfo{author}{\bibfnamefont{K.}~\bibnamefont{Yamamoto}},
  \bibinfo{author}{\bibfnamefont{N.}~\bibnamefont{Irie}},
  \bibinfo{author}{\bibfnamefont{S.}~\bibnamefont{Horita}},
  \bibinfo{author}{\bibfnamefont{T.}~\bibnamefont{Kambe}},
  \bibinfo{author}{\bibfnamefont{N.}~\bibnamefont{Nagao}},
  \bibinfo{author}{\bibfnamefont{K.}~\bibnamefont{Oshima}},
  \bibinfo{author}{\bibfnamefont{N.}~\bibnamefont{Ikeda}}, \bibnamefont{and}
  \bibinfo{author}{\bibfnamefont{T.}~\bibnamefont{Nakamura}},
  \bibinfo{journal}{J. Phys. IV France} \textbf{\bibinfo{volume}{131}},
  \bibinfo{pages}{39} (\bibinfo{year}{2005}).

\bibitem[{\citenamefont{Transtrum et~al.}(2015)\citenamefont{Transtrum, Machta,
  Brown, Daniels, Myers, and Sethna}}]{transtrum15}
\bibinfo{author}{\bibfnamefont{M.~K.} \bibnamefont{Transtrum}},
  \bibinfo{author}{\bibfnamefont{B.~B.} \bibnamefont{Machta}},
  \bibinfo{author}{\bibfnamefont{K.~S.} \bibnamefont{Brown}},
  \bibinfo{author}{\bibfnamefont{B.~C.} \bibnamefont{Daniels}},
  \bibinfo{author}{\bibfnamefont{C.~R.} \bibnamefont{Myers}}, \bibnamefont{and}
  \bibinfo{author}{\bibfnamefont{J.~P.} \bibnamefont{Sethna}},
  \bibinfo{journal}{The Journal of Chemical Physics}
  \textbf{\bibinfo{volume}{143}}, \bibinfo{eid}{010901} (\bibinfo{year}{2015}).

\bibitem[{jes()}]{jeschkecode}
\bibinfo{note}{Tight binding fitting code written by H. O. Jeschke, used with
  permission.}

\bibitem[{\citenamefont{Nakamura et~al.}(2009)\citenamefont{Nakamura,
  Yoshimoto, Kosugi, Arita, and Imada}}]{nakamura09}
\bibinfo{author}{\bibfnamefont{K.}~\bibnamefont{Nakamura}},
  \bibinfo{author}{\bibfnamefont{Y.}~\bibnamefont{Yoshimoto}},
  \bibinfo{author}{\bibfnamefont{T.}~\bibnamefont{Kosugi}},
  \bibinfo{author}{\bibfnamefont{R.}~\bibnamefont{Arita}}, \bibnamefont{and}
  \bibinfo{author}{\bibfnamefont{M.}~\bibnamefont{Imada}},
  \bibinfo{journal}{Journal of the Physical Society of Japan}
  \textbf{\bibinfo{volume}{78}}, \bibinfo{pages}{083710}
  (\bibinfo{year}{2009}).

\bibitem[{\citenamefont{Nakamura et~al.}(2012)\citenamefont{Nakamura,
  Yoshimoto, and Imada}}]{nakamura12}
\bibinfo{author}{\bibfnamefont{K.}~\bibnamefont{Nakamura}},
  \bibinfo{author}{\bibfnamefont{Y.}~\bibnamefont{Yoshimoto}},
  \bibnamefont{and} \bibinfo{author}{\bibfnamefont{M.}~\bibnamefont{Imada}},
  \bibinfo{journal}{Phys. Rev. B} \textbf{\bibinfo{volume}{86}},
  \bibinfo{pages}{205117} (\bibinfo{year}{2012}).

\bibitem[{\citenamefont{Altmeyer et~al.}(2015)\citenamefont{Altmeyer,
  Valent\'{i}, and Jeschke}}]{altmeyer15}
\bibinfo{author}{\bibfnamefont{M.}~\bibnamefont{Altmeyer}},
  \bibinfo{author}{\bibfnamefont{R.}~\bibnamefont{Valent\'{i}}},
  \bibnamefont{and} \bibinfo{author}{\bibfnamefont{H.~O.}
  \bibnamefont{Jeschke}}, \bibinfo{journal}{Phys. Rev. B}
  \textbf{\bibinfo{volume}{91}}, \bibinfo{pages}{245137}
  (\bibinfo{year}{2015}).

\bibitem[{\citenamefont{Jacko et~al.}(2010)\citenamefont{Jacko, Powell, and
  McKenzie}}]{jacko10a}
\bibinfo{author}{\bibfnamefont{A.~C.} \bibnamefont{Jacko}},
  \bibinfo{author}{\bibfnamefont{B.~J.} \bibnamefont{Powell}},
  \bibnamefont{and} \bibinfo{author}{\bibfnamefont{R.~H.}
  \bibnamefont{McKenzie}}, \bibinfo{journal}{J. Chem. Phys.}
  \textbf{\bibinfo{volume}{133}}, \bibinfo{pages}{124314}
  (\bibinfo{year}{2010}).

\bibitem[{\citenamefont{Komatsu et~al.}(1996)\citenamefont{Komatsu, Matsukawa,
  Inoue, and Saito}}]{komatsu96}
\bibinfo{author}{\bibfnamefont{T.}~\bibnamefont{Komatsu}},
  \bibinfo{author}{\bibfnamefont{N.}~\bibnamefont{Matsukawa}},
  \bibinfo{author}{\bibfnamefont{T.}~\bibnamefont{Inoue}}, \bibnamefont{and}
  \bibinfo{author}{\bibfnamefont{G.}~\bibnamefont{Saito}},
  \bibinfo{journal}{Journal of the Physical Society of Japan}
  \textbf{\bibinfo{volume}{65}}, \bibinfo{pages}{1340} (\bibinfo{year}{1996}).

\bibitem[{\citenamefont{McKenzie}(1998)}]{mckenzie98}
\bibinfo{author}{\bibfnamefont{R.~H.} \bibnamefont{McKenzie}},
  \bibinfo{journal}{Comments Cond. Mat. Phys.} \textbf{\bibinfo{volume}{18}},
  \bibinfo{pages}{309} (\bibinfo{year}{1998}).

\bibitem[{\citenamefont{Scriven and Powell}(2009)}]{scriven09b}
\bibinfo{author}{\bibfnamefont{E.}~\bibnamefont{Scriven}} \bibnamefont{and}
  \bibinfo{author}{\bibfnamefont{B.~J.} \bibnamefont{Powell}},
  \bibinfo{journal}{Phys. Rev. B} \textbf{\bibinfo{volume}{80}},
  \bibinfo{pages}{205107} (\bibinfo{year}{2009}).

\bibitem[{\citenamefont{Yoshimi et~al.}(2012)\citenamefont{Yoshimi, Seo,
  Ishibashi, and Brown}}]{yoshimi12}
\bibinfo{author}{\bibfnamefont{K.}~\bibnamefont{Yoshimi}},
  \bibinfo{author}{\bibfnamefont{H.}~\bibnamefont{Seo}},
  \bibinfo{author}{\bibfnamefont{S.}~\bibnamefont{Ishibashi}},
  \bibnamefont{and} \bibinfo{author}{\bibfnamefont{S.~E.} \bibnamefont{Brown}},
  \bibinfo{journal}{Phys. Rev. Lett.} \textbf{\bibinfo{volume}{108}},
  \bibinfo{pages}{096402} (\bibinfo{year}{2012}).

\bibitem[{\citenamefont{Tsumuraya et~al.}(2013)\citenamefont{Tsumuraya, Seo,
  Tsuchiizu, Kato, and Miyazaki}}]{tsumuraya13}
\bibinfo{author}{\bibfnamefont{T.}~\bibnamefont{Tsumuraya}},
  \bibinfo{author}{\bibfnamefont{H.}~\bibnamefont{Seo}},
  \bibinfo{author}{\bibfnamefont{M.}~\bibnamefont{Tsuchiizu}},
  \bibinfo{author}{\bibfnamefont{R.}~\bibnamefont{Kato}}, \bibnamefont{and}
  \bibinfo{author}{\bibfnamefont{T.}~\bibnamefont{Miyazaki}},
  \bibinfo{journal}{Journal of the Physical Society of Japan}
  \textbf{\bibinfo{volume}{82}}, \bibinfo{pages}{033709}
  (\bibinfo{year}{2013}).

\bibitem[{\citenamefont{Nakamura et~al.}(2006)\citenamefont{Nakamura, Arita,
  Yoshimoto, and Tsuneyuki}}]{nakamura06}
\bibinfo{author}{\bibfnamefont{K.}~\bibnamefont{Nakamura}},
  \bibinfo{author}{\bibfnamefont{R.}~\bibnamefont{Arita}},
  \bibinfo{author}{\bibfnamefont{Y.}~\bibnamefont{Yoshimoto}},
  \bibnamefont{and}
  \bibinfo{author}{\bibfnamefont{S.}~\bibnamefont{Tsuneyuki}},
  \bibinfo{journal}{Phys. Rev. B} \textbf{\bibinfo{volume}{74}},
  \bibinfo{pages}{235113} (\bibinfo{year}{2006}).

\bibitem[{\citenamefont{DeVoe}(1964)}]{devoe64}
\bibinfo{author}{\bibfnamefont{H.}~\bibnamefont{DeVoe}}, \bibinfo{journal}{J.
  Chem. Phys.} \textbf{\bibinfo{volume}{41}}, \bibinfo{pages}{393}
  (\bibinfo{year}{1964}).

\bibitem[{\citenamefont{Cano-Cort{\'e}s
  et~al.}(2010)\citenamefont{Cano-Cort{\'e}s, Dolfen, Merino, and
  Koch}}]{cano10b}
\bibinfo{author}{\bibfnamefont{L.}~\bibnamefont{Cano-Cort{\'e}s}},
  \bibinfo{author}{\bibfnamefont{A.}~\bibnamefont{Dolfen}},
  \bibinfo{author}{\bibfnamefont{J.}~\bibnamefont{Merino}}, \bibnamefont{and}
  \bibinfo{author}{\bibfnamefont{E.}~\bibnamefont{Koch}},
  \bibinfo{journal}{Physica B} \textbf{\bibinfo{volume}{405}},
  \bibinfo{pages}{S185} (\bibinfo{year}{2010}).

\bibitem[{\citenamefont{Aryasetiawan et~al.}(2004)\citenamefont{Aryasetiawan,
  Imada, Georges, Kotliar, Biermann, and Lichtenstein}}]{aryasetiawan04}
\bibinfo{author}{\bibfnamefont{F.}~\bibnamefont{Aryasetiawan}},
  \bibinfo{author}{\bibfnamefont{M.}~\bibnamefont{Imada}},
  \bibinfo{author}{\bibfnamefont{A.}~\bibnamefont{Georges}},
  \bibinfo{author}{\bibfnamefont{G.}~\bibnamefont{Kotliar}},
  \bibinfo{author}{\bibfnamefont{S.}~\bibnamefont{Biermann}}, \bibnamefont{and}
  \bibinfo{author}{\bibfnamefont{A.~I.} \bibnamefont{Lichtenstein}},
  \bibinfo{journal}{Phys. Rev. B} \textbf{\bibinfo{volume}{70}},
  \bibinfo{pages}{195104} (\bibinfo{year}{2004}).

\bibitem[{\citenamefont{Bohm and Pines}(1951)}]{bohm51}
\bibinfo{author}{\bibfnamefont{D.}~\bibnamefont{Bohm}} \bibnamefont{and}
  \bibinfo{author}{\bibfnamefont{D.}~\bibnamefont{Pines}},
  \bibinfo{journal}{Phys. Rev.} \textbf{\bibinfo{volume}{82}},
  \bibinfo{pages}{625} (\bibinfo{year}{1951}).

\bibitem[{\citenamefont{Pines and Bohm}(1952)}]{pines52}
\bibinfo{author}{\bibfnamefont{D.}~\bibnamefont{Pines}} \bibnamefont{and}
  \bibinfo{author}{\bibfnamefont{D.}~\bibnamefont{Bohm}},
  \bibinfo{journal}{Phys. Rev.} \textbf{\bibinfo{volume}{85}},
  \bibinfo{pages}{338} (\bibinfo{year}{1952}).

\bibitem[{\citenamefont{Bohm and Pines}(1953)}]{bohm53}
\bibinfo{author}{\bibfnamefont{D.}~\bibnamefont{Bohm}} \bibnamefont{and}
  \bibinfo{author}{\bibfnamefont{D.}~\bibnamefont{Pines}},
  \bibinfo{journal}{Phys. Rev.} \textbf{\bibinfo{volume}{92}},
  \bibinfo{pages}{609} (\bibinfo{year}{1953}).

\bibitem[{\citenamefont{Pines}(1953)}]{pines53}
\bibinfo{author}{\bibfnamefont{D.}~\bibnamefont{Pines}},
  \bibinfo{journal}{Phys. Rev.} \textbf{\bibinfo{volume}{92}},
  \bibinfo{pages}{626} (\bibinfo{year}{1953}).

\bibitem[{\citenamefont{Gell-Mann and Brueckner}(1957)}]{gellmann57}
\bibinfo{author}{\bibfnamefont{M.}~\bibnamefont{Gell-Mann}} \bibnamefont{and}
  \bibinfo{author}{\bibfnamefont{K.~A.} \bibnamefont{Brueckner}},
  \bibinfo{journal}{Phys. Rev.} \textbf{\bibinfo{volume}{106}},
  \bibinfo{pages}{364} (\bibinfo{year}{1957}).

\bibitem[{rpa()}]{rpabubblewiki}
\bibinfo{note}{RPA bubble diagram produced by wikimedia user `Condmatstrel',
  and released under the Creative Commons Attribution 3.0 license.}

\bibitem[{\citenamefont{Miyake et~al.}(2009)\citenamefont{Miyake, Aryasetiawan,
  and Imada}}]{miyake09}
\bibinfo{author}{\bibfnamefont{T.}~\bibnamefont{Miyake}},
  \bibinfo{author}{\bibfnamefont{F.}~\bibnamefont{Aryasetiawan}},
  \bibnamefont{and} \bibinfo{author}{\bibfnamefont{M.}~\bibnamefont{Imada}},
  \bibinfo{journal}{Phys. Rev. B} \textbf{\bibinfo{volume}{80}},
  \bibinfo{pages}{155134} (\bibinfo{year}{2009}).

\bibitem[{\citenamefont{Miyake and Aryasetiawan}(2008)}]{miyake08}
\bibinfo{author}{\bibfnamefont{T.}~\bibnamefont{Miyake}} \bibnamefont{and}
  \bibinfo{author}{\bibfnamefont{F.}~\bibnamefont{Aryasetiawan}},
  \bibinfo{journal}{Phys. Rev. B} \textbf{\bibinfo{volume}{77}},
  \bibinfo{pages}{085122} (\bibinfo{year}{2008}).

\bibitem[{\citenamefont{Nakamura et~al.}(2008)\citenamefont{Nakamura, Arita,
  and Imada}}]{nakamura08}
\bibinfo{author}{\bibfnamefont{K.}~\bibnamefont{Nakamura}},
  \bibinfo{author}{\bibfnamefont{R.}~\bibnamefont{Arita}}, \bibnamefont{and}
  \bibinfo{author}{\bibfnamefont{M.}~\bibnamefont{Imada}},
  \bibinfo{journal}{Journal of the Physical Society of Japan}
  \textbf{\bibinfo{volume}{77}}, \bibinfo{pages}{093711}
  (\bibinfo{year}{2008}).

\bibitem[{\citenamefont{Kobayashi et~al.}(1991)\citenamefont{Kobayashi,
  Kobayashi, Miyamoto, Kato, Clark, and Underhill}}]{kobayashi91}
\bibinfo{author}{\bibfnamefont{A.}~\bibnamefont{Kobayashi}},
  \bibinfo{author}{\bibfnamefont{H.}~\bibnamefont{Kobayashi}},
  \bibinfo{author}{\bibfnamefont{A.}~\bibnamefont{Miyamoto}},
  \bibinfo{author}{\bibfnamefont{R.}~\bibnamefont{Kato}},
  \bibinfo{author}{\bibfnamefont{R.~A.} \bibnamefont{Clark}}, \bibnamefont{and}
  \bibinfo{author}{\bibfnamefont{A.~E.} \bibnamefont{Underhill}},
  \bibinfo{journal}{Chem. Lett.} \textbf{\bibinfo{volume}{20}},
  \bibinfo{pages}{2063 } (\bibinfo{year}{1991}).

\bibitem[{\citenamefont{Seya et~al.}(1995)\citenamefont{Seya, Kobayashi,
  Nakamura, Takahashi, Osako, Kobayashi, Kato, Kobayashi, and Iguchi}}]{seya95}
\bibinfo{author}{\bibfnamefont{K.}~\bibnamefont{Seya}},
  \bibinfo{author}{\bibfnamefont{Y.}~\bibnamefont{Kobayashi}},
  \bibinfo{author}{\bibfnamefont{T.}~\bibnamefont{Nakamura}},
  \bibinfo{author}{\bibfnamefont{T.}~\bibnamefont{Takahashi}},
  \bibinfo{author}{\bibfnamefont{Y.}~\bibnamefont{Osako}},
  \bibinfo{author}{\bibfnamefont{H.}~\bibnamefont{Kobayashi}},
  \bibinfo{author}{\bibfnamefont{R.}~\bibnamefont{Kato}},
  \bibinfo{author}{\bibfnamefont{A.}~\bibnamefont{Kobayashi}},
  \bibnamefont{and} \bibinfo{author}{\bibfnamefont{H.}~\bibnamefont{Iguchi}},
  \bibinfo{journal}{Synth. Met.} \textbf{\bibinfo{volume}{70}},
  \bibinfo{pages}{1043} (\bibinfo{year}{1995}).

\bibitem[{\citenamefont{Tamura and Kato}(2002)}]{tamura02}
\bibinfo{author}{\bibfnamefont{M.}~\bibnamefont{Tamura}} \bibnamefont{and}
  \bibinfo{author}{\bibfnamefont{R.}~\bibnamefont{Kato}}, \bibinfo{journal}{J.
  Phys.: Condens. Matt.} \textbf{\bibinfo{volume}{14}}, \bibinfo{pages}{L729 }
  (\bibinfo{year}{2002}).

\bibitem[{\citenamefont{Itou et~al.}(2008)\citenamefont{Itou, Oyamada, Maegawa,
  Tamura, and Kato}}]{itou08}
\bibinfo{author}{\bibfnamefont{T.}~\bibnamefont{Itou}},
  \bibinfo{author}{\bibfnamefont{A.}~\bibnamefont{Oyamada}},
  \bibinfo{author}{\bibfnamefont{S.}~\bibnamefont{Maegawa}},
  \bibinfo{author}{\bibfnamefont{M.}~\bibnamefont{Tamura}}, \bibnamefont{and}
  \bibinfo{author}{\bibfnamefont{R.}~\bibnamefont{Kato}},
  \bibinfo{journal}{Phys. Rev. B} \textbf{\bibinfo{volume}{77}},
  \bibinfo{pages}{104413} (\bibinfo{year}{2008}).

\bibitem[{\citenamefont{Manna et~al.}(2014)\citenamefont{Manna, de~Souza, Kato,
  and Lang}}]{rudra14}
\bibinfo{author}{\bibfnamefont{R.~S.} \bibnamefont{Manna}},
  \bibinfo{author}{\bibfnamefont{M.}~\bibnamefont{de~Souza}},
  \bibinfo{author}{\bibfnamefont{R.}~\bibnamefont{Kato}}, \bibnamefont{and}
  \bibinfo{author}{\bibfnamefont{M.}~\bibnamefont{Lang}},
  \bibinfo{journal}{Phys. Rev. B} \textbf{\bibinfo{volume}{89}},
  \bibinfo{pages}{045113} (\bibinfo{year}{2014}).

\bibitem[{\citenamefont{Mayaffre et~al.}(1995)\citenamefont{Mayaffre, Wzietek,
  J\'erome, Lenoir, and Batail}}]{mayaffre95}
\bibinfo{author}{\bibfnamefont{H.}~\bibnamefont{Mayaffre}},
  \bibinfo{author}{\bibfnamefont{P.}~\bibnamefont{Wzietek}},
  \bibinfo{author}{\bibfnamefont{D.}~\bibnamefont{J\'erome}},
  \bibinfo{author}{\bibfnamefont{C.}~\bibnamefont{Lenoir}}, \bibnamefont{and}
  \bibinfo{author}{\bibfnamefont{P.}~\bibnamefont{Batail}},
  \bibinfo{journal}{Phys. Rev. Lett.} \textbf{\bibinfo{volume}{75}},
  \bibinfo{pages}{4122} (\bibinfo{year}{1995}).

\bibitem[{\citenamefont{Kino and Fukuyama}(1996)}]{kino96}
\bibinfo{author}{\bibfnamefont{H.}~\bibnamefont{Kino}} \bibnamefont{and}
  \bibinfo{author}{\bibfnamefont{H.}~\bibnamefont{Fukuyama}},
  \bibinfo{journal}{Journal of the Physical Society of Japan}
  \textbf{\bibinfo{volume}{65}}, \bibinfo{pages}{2158} (\bibinfo{year}{1996}).

\bibitem[{\citenamefont{Vorontsov and Vekhter}(2007)}]{vorontsov07}
\bibinfo{author}{\bibfnamefont{A.~B.} \bibnamefont{Vorontsov}}
  \bibnamefont{and} \bibinfo{author}{\bibfnamefont{I.}~\bibnamefont{Vekhter}},
  \bibinfo{journal}{Phys. Rev. B} \textbf{\bibinfo{volume}{75}},
  \bibinfo{pages}{224502} (\bibinfo{year}{2007}).

\bibitem[{\citenamefont{Milbradt et~al.}(2013)\citenamefont{Milbradt, Bardin,
  Truncik, Huttema, Jacko, Burn, Lo, Powell, and Broun}}]{jacko13et}
\bibinfo{author}{\bibfnamefont{S.}~\bibnamefont{Milbradt}},
  \bibinfo{author}{\bibfnamefont{A.~A.} \bibnamefont{Bardin}},
  \bibinfo{author}{\bibfnamefont{C.~J.~S.} \bibnamefont{Truncik}},
  \bibinfo{author}{\bibfnamefont{W.~A.} \bibnamefont{Huttema}},
  \bibinfo{author}{\bibfnamefont{A.~C.} \bibnamefont{Jacko}},
  \bibinfo{author}{\bibfnamefont{P.~L.} \bibnamefont{Burn}},
  \bibinfo{author}{\bibfnamefont{S.-C.} \bibnamefont{Lo}},
  \bibinfo{author}{\bibfnamefont{B.~J.} \bibnamefont{Powell}},
  \bibnamefont{and} \bibinfo{author}{\bibfnamefont{D.~M.} \bibnamefont{Broun}},
  \bibinfo{journal}{Phys. Rev. B} \textbf{\bibinfo{volume}{88}},
  \bibinfo{pages}{064501} (\bibinfo{year}{2013}).

\bibitem[{\citenamefont{Llusar et~al.}(2004)\citenamefont{Llusar, Uriel,
  Vicent, Clemente-Juan, Coronado, G{\'o}mez-Garcı´a, Bra{\"{\i}}da, and
  Canadell}}]{llusar04}
\bibinfo{author}{\bibfnamefont{R.}~\bibnamefont{Llusar}},
  \bibinfo{author}{\bibfnamefont{S.}~\bibnamefont{Uriel}},
  \bibinfo{author}{\bibfnamefont{C.}~\bibnamefont{Vicent}},
  \bibinfo{author}{\bibfnamefont{J.~M.} \bibnamefont{Clemente-Juan}},
  \bibinfo{author}{\bibfnamefont{E.}~\bibnamefont{Coronado}},
  \bibinfo{author}{\bibfnamefont{C.~J.} \bibnamefont{G{\'o}mez-Garcı´a}},
  \bibinfo{author}{\bibfnamefont{B.}~\bibnamefont{Bra{\"{\i}}da}},
  \bibnamefont{and} \bibinfo{author}{\bibfnamefont{E.}~\bibnamefont{Canadell}},
  \bibinfo{journal}{J. Am. Chem. Soc.} \textbf{\bibinfo{volume}{126}},
  \bibinfo{pages}{12076 } (\bibinfo{year}{2004}).

\bibitem[{\citenamefont{Janani et~al.}(2014{\natexlab{a}})\citenamefont{Janani,
  Merino, McCulloch, and Powell}}]{janani14a}
\bibinfo{author}{\bibfnamefont{C.}~\bibnamefont{Janani}},
  \bibinfo{author}{\bibfnamefont{J.}~\bibnamefont{Merino}},
  \bibinfo{author}{\bibfnamefont{I.~P.} \bibnamefont{McCulloch}},
  \bibnamefont{and} \bibinfo{author}{\bibfnamefont{B.~J.}
  \bibnamefont{Powell}}, \bibinfo{journal}{Phys. Rev. Lett.}
  \textbf{\bibinfo{volume}{113}}, \bibinfo{pages}{267204}
  (\bibinfo{year}{2014}{\natexlab{a}}).

\bibitem[{\citenamefont{Janani et~al.}(2014{\natexlab{b}})\citenamefont{Janani,
  Merino, McCulloch, and Powell}}]{janani14b}
\bibinfo{author}{\bibfnamefont{C.}~\bibnamefont{Janani}},
  \bibinfo{author}{\bibfnamefont{J.}~\bibnamefont{Merino}},
  \bibinfo{author}{\bibfnamefont{I.~P.} \bibnamefont{McCulloch}},
  \bibnamefont{and} \bibinfo{author}{\bibfnamefont{B.~J.}
  \bibnamefont{Powell}}, \bibinfo{journal}{Phys. Rev. B}
  \textbf{\bibinfo{volume}{90}}, \bibinfo{pages}{035120}
  (\bibinfo{year}{2014}{\natexlab{b}}).

\bibitem[{\citenamefont{Jacko et~al.}(2015)\citenamefont{Jacko, Janani,
  Koepernik, and Powell}}]{jacko15a}
\bibinfo{author}{\bibfnamefont{A.~C.} \bibnamefont{Jacko}},
  \bibinfo{author}{\bibfnamefont{C.}~\bibnamefont{Janani}},
  \bibinfo{author}{\bibfnamefont{K.}~\bibnamefont{Koepernik}},
  \bibnamefont{and} \bibinfo{author}{\bibfnamefont{B.~J.}
  \bibnamefont{Powell}}, \bibinfo{journal}{Phys. Rev. B}
  \textbf{\bibinfo{volume}{91}}, \bibinfo{pages}{125140}
  (\bibinfo{year}{2015}).

\bibitem[{\citenamefont{Yao and Kivelson}(2007)}]{yao07}
\bibinfo{author}{\bibfnamefont{H.}~\bibnamefont{Yao}} \bibnamefont{and}
  \bibinfo{author}{\bibfnamefont{S.~A.} \bibnamefont{Kivelson}},
  \bibinfo{journal}{Phys. Rev. Lett.} \textbf{\bibinfo{volume}{99}},
  \bibinfo{pages}{247203} (\bibinfo{year}{2007}).

\bibitem[{\citenamefont{Wen et~al.}(2010)\citenamefont{Wen, R\"uegg, Wang, and
  Fiete}}]{wen10}
\bibinfo{author}{\bibfnamefont{J.}~\bibnamefont{Wen}},
  \bibinfo{author}{\bibfnamefont{A.}~\bibnamefont{R\"uegg}},
  \bibinfo{author}{\bibfnamefont{C.-C.~J.} \bibnamefont{Wang}},
  \bibnamefont{and} \bibinfo{author}{\bibfnamefont{G.~A.} \bibnamefont{Fiete}},
  \bibinfo{journal}{Phys. Rev. B} \textbf{\bibinfo{volume}{82}},
  \bibinfo{pages}{075125} (\bibinfo{year}{2010}).

\bibitem[{\citenamefont{R{\"{u}}egg et~al.}(2010)\citenamefont{R{\"{u}}egg,
  Wen, and Fiete}}]{ruegg10}
\bibinfo{author}{\bibfnamefont{A.}~\bibnamefont{R{\"{u}}egg}},
  \bibinfo{author}{\bibfnamefont{J.}~\bibnamefont{Wen}}, \bibnamefont{and}
  \bibinfo{author}{\bibfnamefont{G.~A.} \bibnamefont{Fiete}},
  \bibinfo{journal}{Phys. Rev. B} \textbf{\bibinfo{volume}{81}},
  \bibinfo{pages}{205115} (\bibinfo{year}{2010}).

\bibitem[{\citenamefont{Tikhonov and Feigel'man}(2010)}]{tikhonov10}
\bibinfo{author}{\bibfnamefont{K.~S.} \bibnamefont{Tikhonov}} \bibnamefont{and}
  \bibinfo{author}{\bibfnamefont{M.~V.} \bibnamefont{Feigel'man}},
  \bibinfo{journal}{Phys. Rev. Lett.} \textbf{\bibinfo{volume}{105}},
  \bibinfo{pages}{067207} (\bibinfo{year}{2010}).

\bibitem[{\citenamefont{Yao et~al.}(2007)\citenamefont{Yao, Laumann, Gorshkov,
  Weimer, Jiang, Cirac, Zoller, and Lukin}}]{yao13}
\bibinfo{author}{\bibfnamefont{N.}~\bibnamefont{Yao}},
  \bibinfo{author}{\bibfnamefont{C.}~\bibnamefont{Laumann}},
  \bibinfo{author}{\bibfnamefont{A.}~\bibnamefont{Gorshkov}},
  \bibinfo{author}{\bibfnamefont{H.}~\bibnamefont{Weimer}},
  \bibinfo{author}{\bibfnamefont{L.}~\bibnamefont{Jiang}},
  \bibinfo{author}{\bibfnamefont{J.}~\bibnamefont{Cirac}},
  \bibinfo{author}{\bibfnamefont{P.}~\bibnamefont{Zoller}}, \bibnamefont{and}
  \bibinfo{author}{\bibfnamefont{M.}~\bibnamefont{Lukin}},
  \bibinfo{journal}{Nat. Comms.} \textbf{\bibinfo{volume}{4}},
  \bibinfo{pages}{1585} (\bibinfo{year}{2007}).

\bibitem[{\citenamefont{Bergman et~al.}(2008)\citenamefont{Bergman, Wu, and
  Balents}}]{bergman08}
\bibinfo{author}{\bibfnamefont{D.~L.} \bibnamefont{Bergman}},
  \bibinfo{author}{\bibfnamefont{C.}~\bibnamefont{Wu}}, \bibnamefont{and}
  \bibinfo{author}{\bibfnamefont{L.}~\bibnamefont{Balents}},
  \bibinfo{journal}{Phys. Rev. B} \textbf{\bibinfo{volume}{78}},
  \bibinfo{pages}{125104} (\bibinfo{year}{2008}).


\bibitem[{comm()}]{nourse}
\bibinfo{note}{H. Nourse, in preparation.}

\end{thebibliography}

\end{document}